\documentclass[bibyear]{aa}
\usepackage{graphicx}
\usepackage{txfonts}
\newcommand{\lpup}{\mbox{L$_2$~Pup}}
\newcommand{\TEFF}{\mbox{$T_{\rm eff}$}}
\newcommand{\RSTAR}{\mbox{$R_{\star}$}}
\newcommand{\LOGG}{\mbox{$\log \varg$}}
\newcommand{\MSTAR}{\mbox{$M_{\star}$}}

\newcommand{\LSOL}{\mbox{$L_{\sun}$}}
\newcommand{\RSOL}{\mbox{$R_{\sun}$}}
\newcommand{\MSOL}{\mbox{$M_{\sun}$}}
\newcommand{\MSOLPERYR}{\mbox{$M_{\sun}$~yr$^{-1}$}}
\newcommand{\micron}{\mbox{$\mu$m}}
\newcommand{\KMS}{\mbox{km s$^{-1}$}}
\newcommand{\tauV}{\mbox{$\tau_{V}$}}
\newcommand{\hin}{\mbox{$h_{\rm in}$}}
\usepackage{color}

\begin{document}

\title{
Infrared interferometric imaging of the compact dust disk 
around the AGB star HR3126 with the bipolar Toby Jug Nebula  
\thanks{
Based on AMBER, GRAVITY, NACO, SPHERE, and VISIR 
observations made with the Very Large Telescope 
and Very Large Telescope Interferometer of the European Southern Observatory. 
Program IDs: 096.D-0482, 098.D-0525, 099.D-0493, and 0102.D-0550.
},
\thanks{
Based on observations with AKARI, a JAXA project with the participation of ESA.
},
\thanks{
{\it Herschel} is an ESA space observatory with science instruments provided by European-led Principal Investigator consortia and with important participation from NASA.}
}
%\subtitle{
%}

\author{K.~Ohnaka\inst{1} 
\and
D.~Schertl\inst{2}
\and
K.-H.~Hofmann\inst{2} 
\and
G.~Weigelt\inst{2} 
}

\offprints{K.~Ohnaka}

\institute{
Instituto de Astronom\'{i}a, Universidad Cat\'{o}lica del Norte, 
Avenida Angamos 0610, Antofagasta, Chile\\
\email{k1.ohnaka@gmail.com}
\and
Max-Planck-Institut f\"{u}r Radioastronomie, 
Auf dem H\"{u}gel 69, 53121 Bonn, Germany
}

\date{Received / Accepted }

\abstract
% Context
{}
% Aim
{
The asymptotic giant branch (AGB) star HR3126, associated
with the arcminute-scale bipolar Toby Jug Nebula, 
provides a rare opportunity to study the emergence of bipolar
structures at the end of the AGB phase. 
Our goal is to image the central region of HR3126 with high spatial 
resolution. 
}
% Methods
{
We carried out long-baseline interferometric observations with AMBER
and GRAVITY (2--2.45~\micron) at the Very Large Telescope Interferometer
(VLTI) with spectral resolutions of 1500 and 4500, 
speckle interferometric observations with VLT/NACO (2.24~\micron), 
and imaging with SPHERE-ZIMPOL (0.55~\micron) and VISIR 
(7.9--19.5~\micron). 
}
% Results
{
The images reconstructed in the continuum at 2.1--2.29~\micron\ 
from the AMBER+GRAVITY data reveal the central star surrounded by 
an elliptical ring-like structure
with a semimajor and semiminor axis of 5.3 and 3.5~mas, respectively. 
The ring is interpreted 
as the inner rim of an equatorial dust disk viewed from an inclination 
angle of $\sim$50\degr, and
its axis is approximately aligned with the arcminute-scale bipolar nebula.
The disk is surprisingly compact, with an inner radius of a mere 
3.5~\RSTAR\ (2~au). 
Our 2-D radiative transfer modeling shows that an optically thick flared disk 
with silicate grains as large as $\sim$4~\micron\ can 
simultaneously reproduce the observed continuum images and 
the spectral energy distribution. 
The images reconstructed in the CO first overtone bands reveal 
elongated extended emission around the central star, 
suggesting the oblateness of the star's 
atmosphere or the presence of a CO gas disk inside the dust cavity. 
The object is unresolved with SPHERE-ZIMPOL, NACO, and VISIR. 
}
% Conclusions
{
If the disk formed together with the bipolar nebula, the grain growth 
from sub-micron to a few microns should have taken place 
over the nebula's dynamical age of $\sim$3900~yrs. 
  The non-detection of a companion in the reconstructed images
  implies that either its 2.2 \micron\ brightness is more than
  $\sim$30 times lower
  than that of the red giant or it might have been shredded due to
  binary interaction.
}

   %  END OF ABSTRACT

\titlerunning{Imaging the compact dust disk around the AGB star HR3126
  with the bipolar Toby Jug Nebula}
\authorrunning{Ohnaka et al.}
\maketitle

\begin{figure*}
\begin{center}
%\resizebox{\hsize}{!}{\rotatebox{0}{\includegraphics{hr3126_opt_2mass.eps}}}
%\resizebox{\hsize}{!}{\rotatebox{0}{\includegraphics{38577F1.eps}}}
\resizebox{\hsize}{!}{\rotatebox{0}{\includegraphics{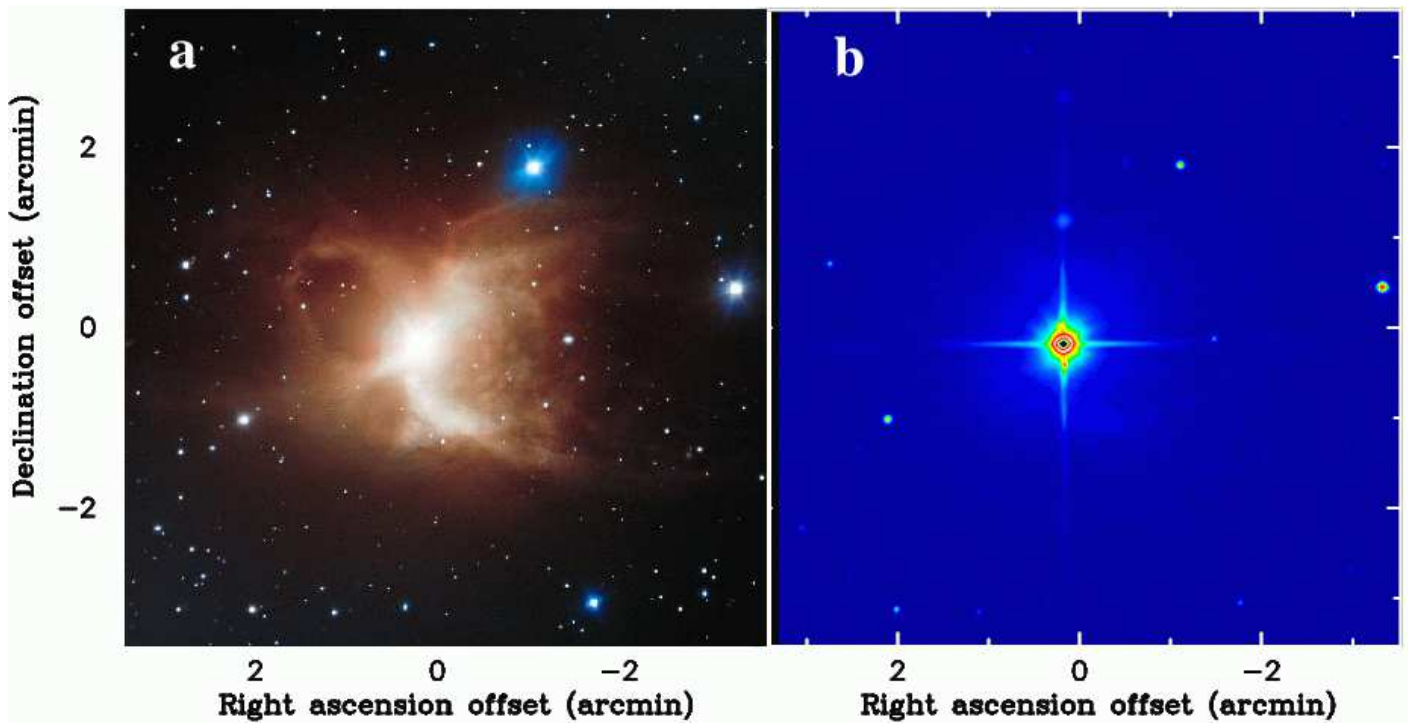}}}
\end{center}
\caption{
  Toby Jug Nebula---the bipolar reflection nebula IC2220
  associated with the AGB star HR3126. 
  {\bf a:} Color-composite optical image taken with VLT/FORS1
  (FOcal Reducer and low dispersion Spectrograph)
  in the ESO Cosmic Gems Programme. 
  The blue shows the $B$-band filter (central wavelength $\lambda_c$ = 429~nm) 
  and [OIII] filter ($\lambda_c$ = 500~nm), the green the $V$-band filter 
  ($\lambda_c$ = 554~nm), the orange the $R$-band filter ($\lambda_c$ = 657~nm),
  and the red the H${\alpha}$ filter ($\lambda_c$ = 656~nm). 
  Credit: ESO.
  {\bf b:} $K_s$-band image from 2MASS (Skrutskie et al. \cite{skrutskie06}).
  The central region of the stellar image is affected by the saturation.
}

\label{hr3126_opt_2mass}
\end{figure*}

\section{Introduction}
\label{sect_intro}

Intermediate- and low-mass stars often develop bipolar or more complex
structures when they evolve from the asymptotic giant
branch (AGB) to protoplanetary nebulae (PPNe) and then to planetary nebulae
(PNe). 
While the physical process responsible for this morphological change
is not yet well understood, binary interaction is considered to play 
an important role (e.g., De Marco \cite{demarco09}). 
Of particular importance is the so-called common-envelope 
evolution, in which the envelope of one of the stars expands and engulfs its 
companion. A number of dynamical physical processes that take place during 
the common-envelope evolution have a significant influence on the 
fate of the binary system. Despite such importance, 
it is very challenging to numerically simulate the common-envelope evolution 
(Ivanova et al. \cite{ivanova13}), and 
it remains unclear how it gives rise to the 
emergence of the bipolar, or more complex structures in PPNe and PNe. 

Observing objects in transition to PPNe and PNe is a direct approach to
improve our understanding of 
the physical process behind the drastic morphological 
change. However, it is difficult to find such objects owing to 
the brevity of the transitional phase. 
A few rare objects show signatures 
of bipolar structures already in the AGB phase, for
example, WX~Psc (Hofmann et al. 
\cite{hofmann01}; Vinkovic et al. \cite{vinkovic04}), 
IRC+10216 (Men'shchikov et al. \cite{menshchikov01}), 
V~Hya (Sahai et al. \cite{sahai09}), $\pi^1$~Gru (Doan et al. \cite{doan17}), 
EP~Aqr (Homan et al. \cite{homan18}; Hoai et al. \cite{hoai19}), 
R~Aqr (Schmid et al. \cite{schmid17}; Melnikov et al. \cite{melnikov18}), 
L$_2$~Pup (Kervella et al. \cite{kervella14}; Ohnaka et al. \cite{ohnaka15}), 
Frosty Leo (Sahai et al. \cite{sahai00};
Lopez et al. \cite{lopez01}; Murakawa et al. \cite{murakawa08}), 
and OH231.8+4.2 (Kastner et al. \cite{kastner98};
Bujarrabal et al. \cite{bujarrabal02};
S\'anchez Contreras et al. \cite{sanchez_contreras18}). 

The target of the present paper, HR3126 (V341~Car, HD65750), 
also exhibits a bipolar structure, 
as can be seen in the optical image shown in Fig.~\ref{hr3126_opt_2mass}a 
taken in the ESO Cosmic Gems 
Programme\footnote{https://www.eso.org/public/images/eso1343a}. 
Of particular note is that 
the central red giant star is associated with a prominent, 
arcminute-scale bipolar optical reflection nebula 
(Dachs \& Isserstedt \cite{dachs73}). 
The nebula is cataloged as IC2220 and also called the Toby Jug Nebula. 
As Fig.~\ref{hr3126_opt_2mass}a shows, 
the angular diameter of the nebula is as large as 
4\arcmin, which corresponds to 0.4~pc at a distance of 381~pc (GAIA 
DR2, \cite{gaia18}, see also Sect.~\ref{sect_basic} 
for discussion about the distance to HR3126). 
The central star is an M1-3II/III star (Dachs \& Isserstedt \cite{dachs73}; 
Humphreys \& Ney \cite{humphreys74}; Chiar et al. \cite{chiar93}), 
an irregular variable with a variability amplitude of $\Delta V \approx 0.9$ 
(Dachs et al. \cite{dachs78}). 
In addition, recent deep imaging with an H$\alpha$ filter has revealed 
two sets of very faint bubble-like nebulosities with diameters of 
10\arcmin\ and 15\arcmin\ (Drudis \cite{drudis18}), both of which are
larger than that of the bipolar reflection nebula. 
These outer nebulosities are suggestive of episodic mass-loss 
events preceding the formation of the bipolar nebula. 
In contrast to the prominent appearance in the visible, the nebula is 
very faint in the near-infrared. 
As the Two Micron All-Sky Survey (2MASS) 
image shows (Fig.~\ref{hr3126_opt_2mass}b), the nebula can barely 
be recognized at 2.2~\micron, where the interferometric imaging presented 
in this paper was carried out. 
The combination of the large-scale bipolar nebula and the central star 
still in the AGB phase (see Sect.~\ref{sect_basic}) provides an interesting 
case for studying the formation of PPNe.

In this paper we present milliarcsecond-resolution near-infrared 
interferometric imaging of the central region of HR3126 using the AMBER 
(Astronomical Multi-BEam combineR) and 
GRAVITY instruments at ESO's Very Large Telescope Interferometer (VLTI). 
We also present speckle interferometric observations taken with VLT/NACO
(Nasmyth Adaptive optics system--COud\'e near-infrared camera), 
polarimetric imaging data taken with VLT/SPHERE-ZIMPOL
(Spectro-Polarimetric High-contrast Exoplanet REsearch
instrument--Zurich IMaging POLarimeter), and mid-infrared 
imaging data taken with VLT/VISIR (VLT Imager and Spectrometer for the
mid-InfraRed). 
We overview the basic properties of HR3126 collected from the literature in 
Sect.~\ref{sect_basic}. 
Then we describe the observations and data 
reduction in Sect.~\ref{sect_obs}. 
The results of the interferometric imaging with AMBER and GRAVITY are 
presented in Sect.~\ref{sect_res}, together with the results 
obtained from SPHERE-ZIMPOL, NACO, and VISIR. 
In Sect.~\ref{sect_modeling}, we present the 2-D radiative transfer modeling
of the observed data, followed by a discussion in Sect.~\ref{sect_discuss}
before closing with the concluding remarks in Sect.~\ref{sect_concl}.

\section{Basic properties of HR3126}
\label{sect_basic}

In this section we discuss the basic properties of HR3126 and its nebula 
derived from previous observations in the literature. 
Castilho et al. (\cite{castilho00}) derived an effective temperature (\TEFF) 
of 3600~K and a surface gravity of \LOGG\ = 0.6 from the analysis of visible 
spectral lines. 
The relationship between the spectral type and effective temperature 
presented by van Belle et al. (\cite{vanbelle99}) shows that 
HR3126's spectral type of M1--3 corresponds to 3700~K to 3800~K. 
We adopted an intermediate value of $3700 \pm 100$~K between the results of
Castilho et al. (\cite{castilho00}) and van Belle et al. (\cite{vanbelle99}). 
The spectral analysis of Castilho et al. (\cite{castilho00}) 
also shows that HR3126 is moderately metal-poor with [Fe/H] = $-0.4$ 
and enriched in the $s$-process elements Zr and Ba with  
[Zr/Fe] = 0.9 and [Ba/Fe] = 0.4. 
These enhancements are similar to those observed in MS-type AGB stars
(Smith \& Lambert \cite{smith85}), suggesting that 
HR3126 is probably in the AGB phase. 
The distance based on the GAIA parallax of $2.623 \pm 0.0931$~mas (GAIA DR2, 
GAIA Collaboration \cite{gaia18}) is $381 \pm 14$~pc.
The distance modulus of 8.0 (distance of 398~pc) derived by Reimers 
(\cite{reimers77}) using the Wilson-Bappu effect of the Ca II K line is
in agreement with the GAIA result. 
We adopted the GAIA distance of 381~pc throughout this paper. 

While the arcminute-scale reflection nebula appears to be very faint 
in the near-infrared, HR3126 exhibits a high infrared excess longward of 
$\sim$2~\micron. It is unusual for an early M giant with a small variability 
amplitude. 
Figure~\ref{hr3126_bestmodel}a shows the observed spectral energy distribution
(SED) of HR3126 based on \mbox{(spectro-)}photometric data available in the
literature: 
photometric data compiled by Chiar et al. (\cite{chiar93}, see also references
therein), measurements with the Diffuse InfraRed Background Experiment 
(DIRBE) instrument on board the COsmic Background Explorer (COBE) 
(Price et al. \cite{price10}), 
spectrum taken with the Short Wavelength Spectrometer (SWS) on board the 
Infrared Space Observatory (ISO) 
(Sloan et al. \cite{sloan03})\footnote{Downloaded from
  https://users.physics.unc.edu/\~{}gcsloan/library/\\swsatlas/atlas.html}, 
and the $N$- and $Q$-band fluxes derived in the present work 
(Sect.~\ref{subsect_res_single}). 
We also included the 9~\micron\ and 18~\micron\ fluxes 
measured with AKARI (Ishihara et al. \cite{ishihara10}) and 
the 70~\micron\ flux measured with the Photodetector Array Camera and
Spectrometer (PACS) instrument
(Poglitsch et al. \cite{poglitsch10}) on board the Herschel Space Observatory
(Pilbratt et al. \cite{pilbratt10}),
taken from the Herschel PACS Point Source Catalog
(European Space Agency \cite{esa17}). 
The interstellar extinction is estimated to be $A_V = 0.223$ or 
$E(B-V) = 0.096$ with $R_v = A_v/E(B-V) = 3.1$ based on 
the 3-D interstellar extinction map published by Lallement et
al. (\cite{lallement19})\footnote{https://astro.acri-st.fr/gaia\_dev}. 
However, Dachs \& Isserstedt (\cite{dachs73}) note that 
the color excess $E(B-V) = 0.38$ observed toward the central star is 
much larger, and therefore, a large fraction of the extinction is 
caused within the nebula.
To obtain the SED from the central region of the object, 
we de-reddened the \mbox{(spectro-)}photometric 
data with $E(B-V) = 0.38$ (or $A_V = 1.18$ with $R_V = 3.1$), 
using the wavelength dependence of the interstellar extinction derived 
by Cardelli et al. (\cite{cardelli89}). 

The radiation of the central star is represented by the photospheric 
model with the parameters closest 
to those of HR3126 (\TEFF\ = 3750~K, \LOGG\ = 0.5, and [Fe/H] = $-0.5$) 
taken from the model grid of Castelli \& Kurucz
(\cite{castelli03})\footnote{http://www.oact.inaf.it/castelli/castelli/grids.html}.
By fitting the photospheric model to the visible and near-infrared 
wavelength regions of the de-reddened observed SED, 
we obtained an angular diameter of $3.0 \pm 0.5$~mas of the central star. 
This corresponds to a linear radius of $122 \pm 20$~\RSOL\
(= $0.57\pm 0.09$~au) at the distance of 381~pc. 
The luminosity is derived to be 
$2500 \pm 860$~\LSOL\ using the linear radius and the effective temperature of 
$3700 \pm 100$~K.
By combining the linear radius of $122 \pm 20$~\RSOL\ with
the surface gravity of
\LOGG\ = 0.6, the current mass is estimated to be $2.2 \pm 0.7$~\MSOL. 
The basic parameters of HR3126 are summarized in Table.~\ref{basic_table}.

The origin of the bipolar reflection nebula is still controversial.
Pesce et al. (\cite{pesce88}) note that the absence of UV emission
at 1150--2000~\AA\ in the International Ultraviolet Explorer (IUE) data
indicates that the central star is not in an interacting binary such as 
a symbiotic star. They also report that 
the Mg II emission at 2800~\AA\ and the UV continuum
longward of $\sim$2700~\AA\ originate from the chromosphere 
typical of an early M giant. 
Based on the analysis of the circumstellar absorption lines in the violet 
spectrum, Reimers (\cite{reimers77}) derived a current mass-loss 
rate of $(2-4) \times 10^{-7}$~\MSOLPERYR\ from the central M giant star 
with an outflow velocity of 13~\KMS. 
However, 
the radio CO line observations by Nyman et al. (\cite{nyman93}) show 
a faster, approximately bipolar outflow at $\sim$35~\KMS\ on a spatial scale 
of the reflection nebula.
The dynamical age of the bipolar outflow in the nebula is then estimated to
be $\sim$3900~yrs based on the linear radius of the nebula ($\sim$0.2~pc) and
the deprojected outflow velocity of 54~\KMS\ (see 
Sect.~\ref{subsect_res_vlti}).

The nebular mass estimated from dust
extinction or scattering in the literature appears to be uncertain. 
Reimers (\cite{reimers77}) estimated the total nebular mass 
to be on the order of $\ga$1~\MSOL\ from the extinction in the nebula. 
Perkins et al. (\cite{perkins81}) derived much smaller values of 
a few $\times 10^{-2}$~\MSOL\ from their polarization observations 
of the nebula. 
  On the other hand, 
  a detailed modeling of the aforementioned radio CO data 
  suggests a total nebular mass of $\sim$1~\MSOL\
  (L.-\AA. Nyman, priv. comm.). 
  This nebular mass and the dynamical age of 3900~yrs mean
  a mass-loss rate of $\sim \! 2.5 \times 10^{-4}$~\MSOLPERYR,
  which is 600--1000 times higher than the current mass-loss rate from
  the central M giant. 
Perhaps the nebular material was ejected in a bipolar manner in a high 
mass-loss event related to the common-envelope evolution, although there is 
no definitive evidence of a binary companion for HR3126 in the literature.
If binary interaction was responsible for the formation of the nebula, 
it is possible that the duration of the high mass-loss
event was much shorter than the dynamical age of $\sim$3900~yrs.
In this case, the mass-loss rate would be even higher than
the $\sim \! 2.5 \times 10^{-4}$~\MSOLPERYR.

The nebular mass of $\sim$1~\MSOL\ and the current stellar mass of
$\sim$2~\MSOL\ 
suggest an initial stellar mass of $\sim$3~\MSOL, if we assume that 
the nebular mass was entirely ejected by the AGB star. 
This is a lower limit because the amount of mass
lost before the formation of the nebula, as revealed by the faint
H$\alpha$ nebulosities, is not clear. 
However, as mentioned above, if binary interaction and significant 
mass transfer were involved in the evolution of HR3126, 
the material in the nebula may originate not only from
the AGB star but also from its companion. In this case, 
it is impossible to estimate the initial mass of two stars. 
Comparison of the observationally derived position of HR3126 on the
Hertzsprung-Russell diagram 
with the theoretical evolutionary tracks of Lagarde et al. (\cite{lagarde12}) 
also suggests an initial mass of $\sim$3~\MSOL. However, given that 
it is questionable whether we can assume single star evolution for HR3126
or not, 
this agreement of the initial mass may be merely fortuitous.

\begin{table}
\caption {
Basic parameters of the AGB star HR3126.
}
\begin{center}
\begin{tabular}{l c l}\hline
Parameter & Value & References \\
\hline
Distance (pc) & $381 \pm 14$ & 1 \\
\TEFF\ (K)    & $3700 \pm 100$  & 2, 3 \\
$L_{\star}$ (\LSOL) &  $2500 \pm 860$   &  4 \\
\LOGG\ (cm~s$^{-2}$)      & 0.6         & 2\\
\RSTAR (\RSOL)  & $122 \pm 20$ & 4 \\
                & ($0.57\pm 0.09$~au) & \\
Angular diameter (mas) & $3.0\pm 0.5$    & 4 \\
Current stellar mass (\MSOL)    &  $2.2 \pm 0.7$  & 4\\
$[{\rm Fe/H}]$      & $-0.4$     &  2 \\
\hline
\label{basic_table}
\vspace*{-7mm}

\end{tabular}
\end{center}
\tablefoot{
  1: GAIA DR2 (\cite{gaia18}). 2: Castilho et al. (\cite{castilho00}).
  3: van Belle et al. (\cite{vanbelle99}). 4: This work. 
}
\end{table}

\begin{figure}
\begin{center}
%\resizebox{\hsize}{!}{\rotatebox{0}{\includegraphics{hr3126_uvplot.ps}}}
\resizebox{\hsize}{!}{\rotatebox{0}{\includegraphics{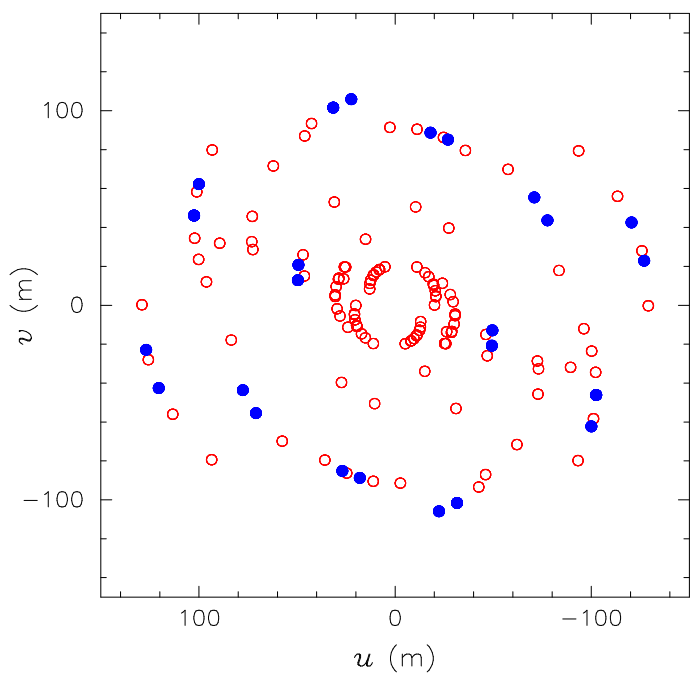}}}
\end{center}
\caption{
$u\varv$ coverage of our VLTI/AMBER (red open circles) and GRAVITY 
(blue filled circles) observations of HR3126.
}
\label{hr3126_uvplot}
\end{figure}

\begin{figure}
\begin{center}
%\resizebox{\hsize}{!}{\rotatebox{0}{\includegraphics{hr3126_spec.ps}}}
\resizebox{\hsize}{!}{\rotatebox{0}{\includegraphics{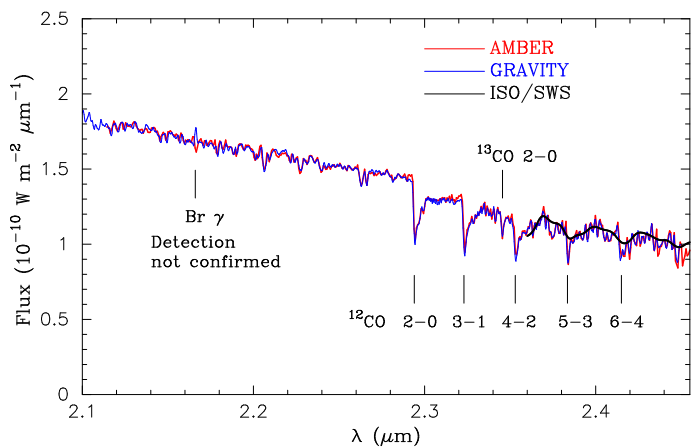}}}
\end{center}
\caption{
Spectra of HR3126 derived from our AMBER (red) and GRAVITY (blue) 
observations, together with the ISO/SWS spectrum (black, Sloan et al.
\cite{sloan03}). 
The CO band heads are marked with the ticks. 
The position of the Br$\gamma$ line, which is not confirmed in the 
spectra of HR3126, is also shown. 
}
\label{hr3126_spec}
\end{figure}

\section{Observations and data reduction}
\label{sect_obs}

\subsection{Near-infrared interferometric observations with AMBER and GRAVITY}
\label{subsect_obs_vlti}

We observed HR3126 with the near-infrared spectro-interferometric 
instruments 
AMBER (Petrov et al. \cite{petrov07}) and GRAVITY (GRAVITY Collaboration 
\cite{gravity17}), as summarized in Table~\ref{obs_vlti_log}. 
Our AMBER observations were carried out in the wavelength region between 2.1 
and 2.45~\micron\ with a spectral resolution of 1500, using five different 
Auxiliary Telescope (AT) triplets (Program ID: 098.D-0525, P.I.: K.~Ohnaka). 
As the $u\varv$ coverage plotted in Fig.~\ref{hr3126_uvplot} shows, 
the projected baseline lengths range from 16~m to 129~m. 
Spectrally dispersed interferograms were recorded with a detector 
integration time (DIT) of 200~ms, using the VLTI fringe tracker FINITO
(Le Bouquin et al. \cite{lebouquin08}). 
We observed $\beta$~Pyx (G5II/III, uniform-disk diameter = $1.9 \pm 0.18$~mas, 
Bourges et al. \cite{bourges17}) for interferometric and spectroscopic 
calibration.
Each AMBER measurement was carried out in a calibrator--science--calibrator
sequence. 

The AMBER data were reduced with amdlib 
ver~3.0.8\footnote{http://www.jmmc.fr/data\_processing\_amber.htm} 
(\mbox{Tatulli} et al. \cite{tatulli07}; Chelli et al. \cite{chelli09}). 
The interferometric observables (squared visibility amplitude, differential 
phase, and closure phase) were extracted by discarding 20\% of the 
frames with the lowest fringe S/N. 
The use of the fringe tracker FINITO can introduce systematic errors in the 
absolute visibility calibration, if the fringe tracking performance is 
significantly different for the science target and the calibrator. 
To minimize this effect, we computed the standard deviation of the phase 
measured with FINITO from the VLTI Reflective Memory Network Recorder (RMNrec) 
data (Le Bouquin et al. \cite{lebouquin09}; M\'erand et al. \cite{merand12}). 
We calibrated the data of HR3126 with those of $\beta$~Pyx only when the 
difference in the FINITO phase standard deviation is within a factor of 
1.2.
The wavelength calibration was done 
using the telluric lines identified in the observed spectrum of the 
calibrator $\beta$~Pyx. The uncertainty in the wavelength calibration 
is $2.0 \times 10^{-4}$~\micron.

Our GRAVITY observations (Program ID: 0102.D-0550, P.I.: K.~Ohnaka)
were carried out in the $K$ band (1.99--2.45~\micron) 
with a spectral resolution of 4500 in the combined polarization mode. 
HR3126 was observed in the single-field mode, using the most extended 
A0-G1-J2-J3 AT quadruplet, 
which provided projected baseline lengths from 51~m to 129~m. 
GRAVITY is always operated with its own fringe tracker. 
We observed $\iota$~Car (A7Ib, uniform-disk diameter = $1.8 \pm 0.19$~mas, 
Bourges et al. \cite{bourges17}) as the calibrator. 
Two data sets were taken on 2018 December 19 (UTC) in a
calibrator--science--calibrator--science--calibrator
sequence.   
The obtained data were reduced with the GRAVITY pipeline 
ver.~1.1.2\footnote{https://www.eso.org/sci/software/pipelines/gravity/}. 
Each data set of HR3126 was calibrated with three data sets of the
calibrator. 
The wavelength calibration was done using the telluric lines identified 
in the observed spectrum of the calibrator as in the case of the AMBER data.

The wavelength ranges covered with AMBER and GRAVITY include the CO first 
overtone bands. 
To reconstruct images in the CO bands by combining the 
AMBER and GRAVITY data, it is necessary to have interferometric data obtained 
with approximately the same spectral resolution. Because the difference 
in the spectral resolution between the AMBER and GRAVITY data is a 
factor of three, we spectrally binned the raw GRAVITY data (both the 
science target and the calibrator) with a running 
box car filter to match the spectral resolution of the AMBER data. 
The spectrally binned data were reduced with the GRAVITY pipeline and 
calibrated in the same manner as the original data. 

Several AMBER data points were taken at approximately the same $u\varv$ 
points but several months apart. Similarly, there are a few data points 
taken with AMBER and GRAVITY at nearly the same $u\varv$ points taken 
1.8 years apart (see Fig.~\ref{hr3126_vis_amber_gravity}). 
These data show no significant time variations. 
The consistency between the AMBER and GRAVITY data shows that the 
calibration of both data is reliable. 
These results justify the 
reconstruction of images combining all AMBER and GRAVITY data.

The spectroscopic calibration to remove the telluric lines and 
instrumental effects was carried out as described in
Appendix~\ref{appendix_spec}. 
Figure~\ref{hr3126_spec} shows the spectroscopically calibrated 
spectra of HR3126 obtained from the AMBER and GRAVITY data (these latter data 
were convolved to the AMBER's spectral resolution as mentioned above). 
Both spectra agree very well, demonstrating the reliability of our 
spectroscopic calibration method. 
The $^{12}$CO first overtone bands from $\varv$ = 2 -- 0 
to 6 -- 4 as well as the $^{13}$CO $\varv$ = 2 -- 0 band can be identified. 
We note that the AMBER spectrum shows very weak absorption at the wavelength
of the Br$\gamma$ line, while the GRAVITY spectrum shows weak emission. 
This is because the proxy stars used to approximate the calibrators' spectra
(see Appendix~\ref{appendix_spec}) show the Br$\gamma$ line in 
absorption, and the slight difference between the proxy stars and the 
calibrators actually observed leads to the weak, residual absorption or 
emission.
The present data set an upper limit of 5\% of the continuum 
on the absorption or emission of the Br$\gamma$ line but do not allow us
to draw a definitive conclusion about the presence
of the Br$\gamma$ line in HR3126. 
However, even if it is absent, it is not unexpected for an early M giant.
We also confirmed that no signature of the Br$\gamma$ line is seen
in the visibility amplitude, closure phase, and differential phase
(even in the original GRAVITY data with the spectral resolution of 4500).

\subsection{Single-dish observations with VLT/SPHERE-ZIMPOL,
NACO, and VISIR}
\label{subsect_obs_single}

We carried out complementary single-dish high spatial resolution 
observations of HR3126 at 0.55, 2.24, and 7.9--19.5~\micron, using the
VLT instruments SPHERE-ZIMPOL (Beuzit et al. \cite{beuzit08};
Thalmann et al. \cite{thalmann08}), NACO (Lenzen et al. \cite{lenzen03}; 
\mbox{Rousset} et al. \cite{rousset03}),
and VISIR (Lagage et al. \cite{lagage04}),
respectively (Program IDs: 096.D-0482 and 099.D-0493, P.I.: K.~Ohnaka). 
The goal of these
single-dish observations is to examine the possible presence of a
component larger than the field of view of the VLTI/AMBER and GRAVITY
observations (250~mas).
The SPHERE-ZIMPOL instrument was used to take nearly diffraction-limited
polarimetric images at 0.55~\micron\ with a spatial resolution of 25~mas 
and a field of view was 2\arcsec $\times$ 2\arcsec. 
We also carried out speckle interferometric observations with NACO at
2.24~\micron, which falls onto the continuum
region in the spectral windows of the AMBER and GRAVITY observations. 
These NACO data provided the visibility amplitude and closure phase 
with a spatial resolution of 64~mas and 
a field of view of 3.5\arcsec$\times$3.5\arcsec. 
Our VISIR observations were made at 7.9, 9.8, 12.8, 17.7, and 19.5~\micron\
with spatial resolutions of 0\farcs4--0\farcs6 and a field of view of
38\arcsec $\times$ 38\arcsec. 
The details of these VLT single-dish observations and data reduction are
described in Appendix~\ref{appendix_obs_single}.

\begin{figure*}
\begin{center}
\resizebox{\hsize}{!}{\rotatebox{-90}{\includegraphics{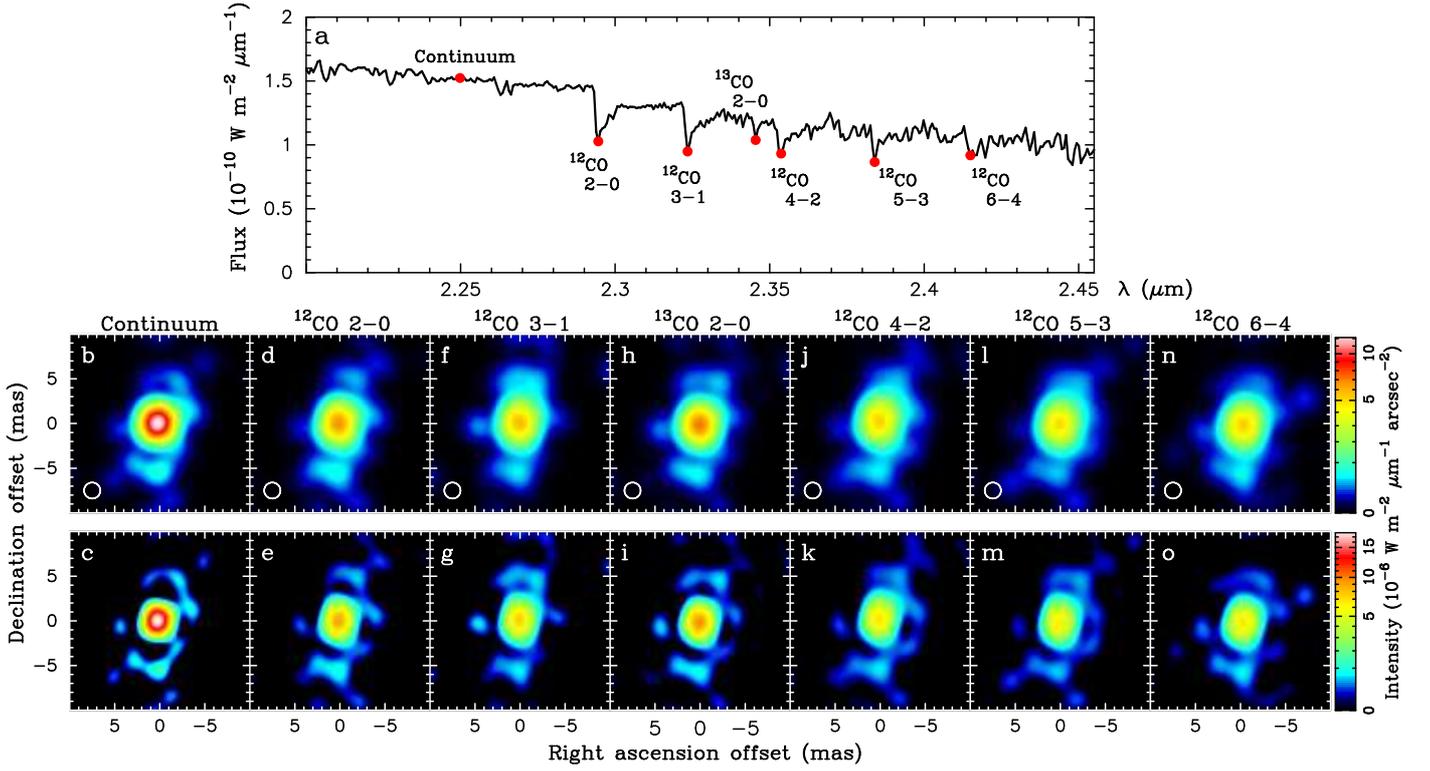}}}
%\resizebox{\hsize}{!}{\rotatebox{-90}{\includegraphics{hr3126_images_qsmooth_7x2_sqrtcolorbar.ps}}}
\end{center}
\caption{
  Wavelength-dependent images of the central region of the Toby Jug Nebula
  around the AGB star HR3126. 
The images reconstructed in the continuum and six $^{12}$CO and $^{13}$CO 
first overtone band heads are shown. 
{\bf a:} Observed spectrum of HR3126. The images reconstructed at the 
wavelength channels marked with the filled dots are shown in the panels 
below. 
{\bf b}--{\bf o:} Columns correspond to the wavelength channel in the 
continuum or in the CO bands as marked in panel {\bf a}. 
The upper and lower rows show the images convolved with the 
beam size of 1.8~mas (shown in the lower left corner of each panel) 
and the unconvolved images, respectively. 
The images are flux-calibrated as described in Sect.~\ref{subsect_res_vlti}. 
}
\label{hr3126_images}
\end{figure*}

\begin{figure*}
\begin{center}
\resizebox{\hsize}{!}{\rotatebox{0}{\includegraphics{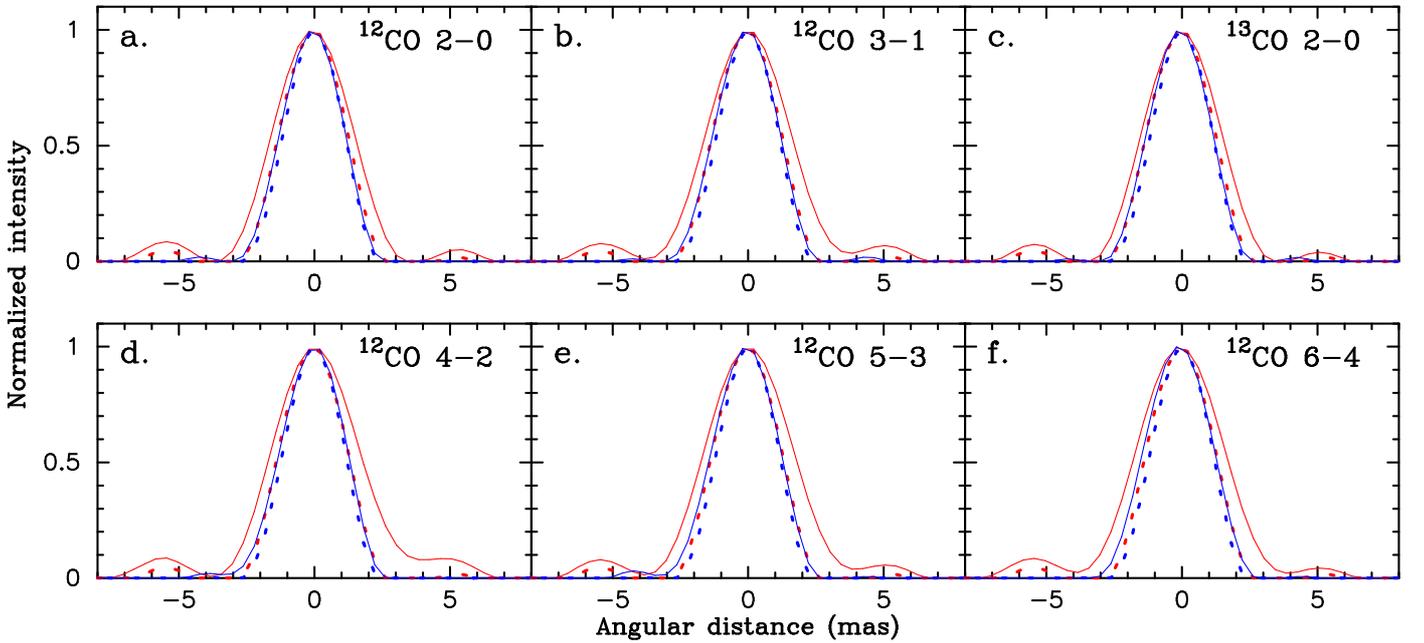}}}
%\resizebox{\hsize}{!}{\rotatebox{0}{\includegraphics{hr3126_1dcuts_qsmooth.ps}}}
\end{center}
\caption{
  Intensity profiles of the (unconvolved) reconstructed images of HR3126
  in the CO band
  heads at two orthogonal position angles. In each panel, the solid red and blue
  lines represent the intensity profiles in the corresponding CO
  band head at position angles of 170\degr\ (approximately N-S direction)
  and 80\degr\ (roughly E-W direction),
  respectively. The dotted red and blue lines represent the intensity
  profiles at 2.25~\micron\ in the continuum at position angles of
  170\degr\ and 80\degr, respectively, and they are the same in all panels.
}
\label{hr3126_1dcuts_qsmooth}
\end{figure*}

\section{Results}
\label{sect_res}

\subsection{AMBER+GRAVITY near-infrared interferometric imaging}
\label{subsect_res_vlti}

We reconstructed images from the AMBER and GRAVITY data (squared visibility 
amplitude and closure phase), using 
MiRA ver~0.0.9\footnote{https://cral-perso.univ-lyon1.fr/labo/perso/eric.thiebaut\\/?Software/MiRA} (Thi\'ebaut \cite{thiebaut08}). 
We employed two different regularization schemes: pixel intensity quadratic 
regularization 
(e.g., Eq. 44 of Thi\'ebaut \cite{thiebaut08},
Eq. 7 of Hofmann et al. \cite{hofmann14},
also called Tikhonov regularization) and the pixel difference quadratic
regularization 
(Eq. 46 of Thi\'ebaut \cite{thiebaut08}, Eq. 9 of
Hofmann et al. \cite{hofmann14}, also called total squared variation
in Event Horizon Telescope Collaboration et al. \cite{eht19}). 
In the case of the pixel intensity quadratic regularization, a flat 
prior was used, while no prior was used for the pixel difference quadratic 
regularization. 
The optimal value of the hyper-parameter $\mu$, 
which represents the weight of the regularization term, was determined 
with the L-curve method (Hansen \cite{hansen92}). The L-curves obtained
with two regularizations for the reconstruction of the continuum images 
are shown in Fig.~\ref{l-curves}.
We checked the L-curve separately for the continuum and each CO band head
and confirmed that the best $\mu$ value for the continuum is also optimal
for the reconstruction of the CO band images.

The reconstruction was started from a point source with each regularization. 
The images reconstructed with two different regularizations 
turned out to be similar. Therefore, we only present below 
the images reconstructed with the pixel difference quadratic regularization. 
The images reconstructed with the pixel intensity
quadratic regularization are shown in Fig.~\ref{hr3126_images_quadratic}. 

The output of MiRA is unconvolved images that minimize the
cost function consisting of the data term (i.e., goodness of the fit
to the interferometric measurements) and the regularization
term. 
In addition to these unconvolved images, we present below the images
convolved with the beam size of $\lambda/2B_{\rm max}$ = 1.8~mas
($B_{\rm max}$ is the maximum baseline),
as is often presented in the literature (e.g., Kraus et al. \cite{kraus10}).
We note that MiRA searches for a solution by comparing the Fourier
transform of unconvolved images with measured interferometric observables,
and therefore, the convolution with the above beam size does not affect
the image reconstruction process. 
The reconstructed images were flux-calibrated so that the flux integrated 
over the entire image matches the photometrically calibrated flux at the 
corresponding wavelength, using the spectrum obtained in 
Sect.~\ref{subsect_obs_vlti}.

Figure~\ref{hr3126_images} shows the images reconstructed 
at seven different wavelengths in the continuum as well as in the CO 
band heads. 
Comparison of the observed visibilities and closure phases with those from 
the reconstructed images is shown in Fig.~\ref{fit_interf}. 
We show both the unconvolved images and the images convolved
with $\lambda/2B_{\rm max}$ = 1.8~mas, as mentioned above. 
The images reconstructed in the continuum 
(Figs.~\ref{hr3126_images}b and \ref{hr3126_images}c) reveal strong central 
emission and an elliptical ring-like structure,
whose eastern side is incomplete. 
The semimajor and semiminor axes of the ring are 5.3~mas and 3.5~mas, 
respectively.

The elliptical ring is interpreted as the inner rim of an equatorial dust disk. 
If we assume that the (deprojected) disk's inner rim is circular, the 
observed ratio of the major and minor axes suggests an inclination angle 
of $\sim$50\degr\ (i.e., we are looking at the central star at $\sim$50\degr\ 
away from pole-on). 
The position angle of the ring's minor axis is 77\degr, 
which means that the disk's symmetry axis is approximately aligned with 
the arcminute-scale bipolar nebula shown in Fig.~\ref{hr3126_opt_2mass}a. 
The outflow velocity of the bipolar nebula of 35~\KMS\ (Nyman et al.
\cite{nyman93}) translates into 
a deprojected velocity of 54~\KMS\ with the inclination angle of 
50\degr.
This deprojected outflow velocity and the angular radius of the bipolar
nebula of 0.2~pc give the dynamical age of 3900~yrs as mentioned in
Sect.~\ref{sect_basic}. 

We interpret the central emission as the central star. 
The emission appears to be smooth (see also Fig. \ref{hr3126_1dcuts_qsmooth}),
not showing the edge of the stellar limb. 
This is simply because our AMBER+GRAVITY measurements sample only the first
visibility lobe of the central star component and therefore do not allow us
to reconstruct the edge of the stellar disk. 
As described in Sect.~\ref{sect_basic}, the angular radius of the central 
star is estimated to be 1.5~mas (or 0.57~au). 
Therefore, the disk is surprisingly compact---in spite of the arcminute-scale
bipolar nebula---with the inner rim being located at a mere 3.5~\RSTAR 
(= 2~au) from the central star.

Figure~\ref{hr3126_images} also shows the images reconstructed 
in the CO band heads of the $\varv$ = 2 -- 0, 3 -- 1, 4 -- 2, 5 -- 3, 
6 -- 4 transitions as well as in the $^{13}$CO 2 -- 0 band head. 
The CO band head images show the inner rim of the equatorial disk 
and the central emission as found in the continuum image described above. 
The intensity of the disk's inner rim in the CO band head images 
is approximately the same as in the continuum image. This is 
expected because the dust opacity changes little across the CO band heads. 
The peak intensity of the central emission in the CO band head images 
is lower than in the continuum image. This can be attributed to the 
CO absorption due to the central star's atmosphere. 

The central emission seen in all CO band head images is more 
extended in the N-S direction than in the continuum image. 
This can be clearly seen in Fig.~\ref{hr3126_1dcuts_qsmooth}, which shows 
the 1-D intensity profiles at two orthogonal
position angles of 170\degr\ and 80\degr\ derived from the unconvolved
images. The 1-D intensity profiles at the
position angle of 170\degr\ at all CO band heads (solid red lines) 
are more extended than that in the continuum (dotted red lines).
The 1-D profiles at the position angle of 80\degr\ only show small
differences between the CO band heads (solid blue lines) and the
continuum (dotted blue lines). 
This suggests that either the central star's atmosphere is more extended 
in the equatorial plane or the CO gas emission from the equatorial plane 
inside the dust cavity makes the central emission appear to be elongated. 
The former scenario is possible if there is an undetected companion 
close to the star. 
Unfortunately, the presence of a companion is neither 
confirmed nor excluded for HR3126, owing to the lack of monitoring 
measurements of the radial velocity
(see discussion in Sect.~\ref{subsect_companion}). 
The two-epoch measurements by Andersen et al. (\cite{andersen85}) 
detected no significant radial velocity variation, although it may simply be a 
coincidence. 
At the moment, we cannot distinguish two scenarios.

\subsection{VLT single-dish imaging and interferometric observations}
\label{subsect_res_single}

HR3126 is unresolved in all our single-dish observations with
SPHERE-ZIMPOL, NACO, and VISIR. It is not surprising that the compact
dust disk of $10.6 \times 7.0$~mas imaged with AMBER and GRAVITY 
is not resolved even with the 25~mas resolution of SPHERE-ZIMPOL.
However, our goal of these single-dish observations was to
examine the possible presence of an extended component.
The result that
HR3126 is unresolved with these instruments suggests the absence of an
extended component that would have been resolved out with AMBER and GRAVITY.
The arcminute-scale bipolar nebula is not seen in the SPHERE-ZIMPOL data, 
probably because its surface brightness is very low 
(Dachs \& Isserstedt \cite{dachs73}; Dachs et al. \cite{dachs78}). 
Details of the results of our VLT single-dish observations are presented in
Appendix~\ref{appendix_res_single}.

Figure~\ref{hr3126_midir_spec} shows the $N$- and $Q$-band fluxes measured 
with VISIR (values are listed in Table~\ref{res_visir_table}), 
together with the ISO/SWS spectrum (Sloan et al. \cite{sloan03})
and AKARI data (Ishihara et al. \cite{ishihara10}). 
The 1\farcs5-radius aperture used to derive the fluxes from the VISIR data
is much smaller than 
the field of view of the ISO/SWS data, 14\arcsec\ $\times$ (20--27)\arcsec, 
and the beam size of the AKARI data, $\sim$9\arcsec\
(Ishihara et al. \cite{ishihara10}). 
Nevertheless, the VISIR fluxes are only slightly 
lower than or equal to the ISO/SWS spectrum and AKARI fluxes.
Therefore, most of the mid-infrared emission from 
7.9 to 19.5~\micron\ is concentrated in the region unresolved with VISIR, 
smaller than the full width at half maximum (FWHM)
of $\sim$0\farcs6 (or half width at half maximum of $\sim$0\farcs3) 
of the observed images. 
This angular size translates into a linear radius of $\sim$110~au or 
$\sim$200~\RSTAR. 

We note that the mid-infrared spectrum of HR3126 only shows a weak, broad 
feature centered at 9--10~\micron, in marked contrast to the prominent 
silicate peak at 9.8~\micron\ often seen in oxygen-rich AGB stars. 
As we show in Sect.~\ref{sect_modeling}, the observed weak feature can be 
explained by the predominance of large grains.

\begin{table}
\caption {
Flux and FWHM of HR3126 measured in the mid-infrared with VLT/VISIR, 
together with FWHM measured on the PSF reference star. 
}
\begin{center}
\begin{tabular}{l c c c c}\hline
Filter & $\lambda_c$  & Flux & FWHM & PSF FWHM \\
       & (\micron)    & (Jy) &(\arcsec) &  (\arcsec) \\
\hline
J7.9  & 7.9  & $102.7\pm 2.1$ & 0.53 & 0.61 \\
SIV\_1 & 9.8 & $98.9\pm 1.5$  & 0.46 & 0.48 \\
NeII  & 12.8 & $61.8\pm 1.1$  & 0.47 & 0.41 \\
Q1    & 17.65 & $50.1\pm 0.9$ & 0.56 & 0.54 \\
Q3    & 19.50 & $38.7\pm 3.1$ & 0.59 & 0.57 \\
\hline
\label{res_visir_table}
\vspace*{-7mm}

\end{tabular}
\end{center}
\end{table}

\begin{figure}
\begin{center}
\resizebox{\hsize}{!}{\rotatebox{0}{\includegraphics{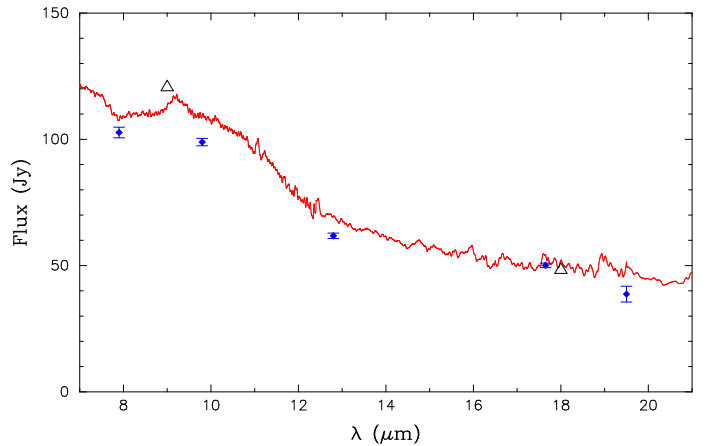}}}
%\resizebox{\hsize}{!}{\rotatebox{0}{\includegraphics{hr3126_midir_spec.ps}}}
\end{center}
\caption{
Mid-infrared observations of HR3126.
The diamonds with the error bars represent our VISIR measurements, 
while the solid line represents the ISO/SWS spectrum.
The AKARI measurements are plotted with the open triangles. 
}
\label{hr3126_midir_spec}
\end{figure}

\begin{figure*}
\resizebox{\hsize}{!}{\rotatebox{0}{\includegraphics{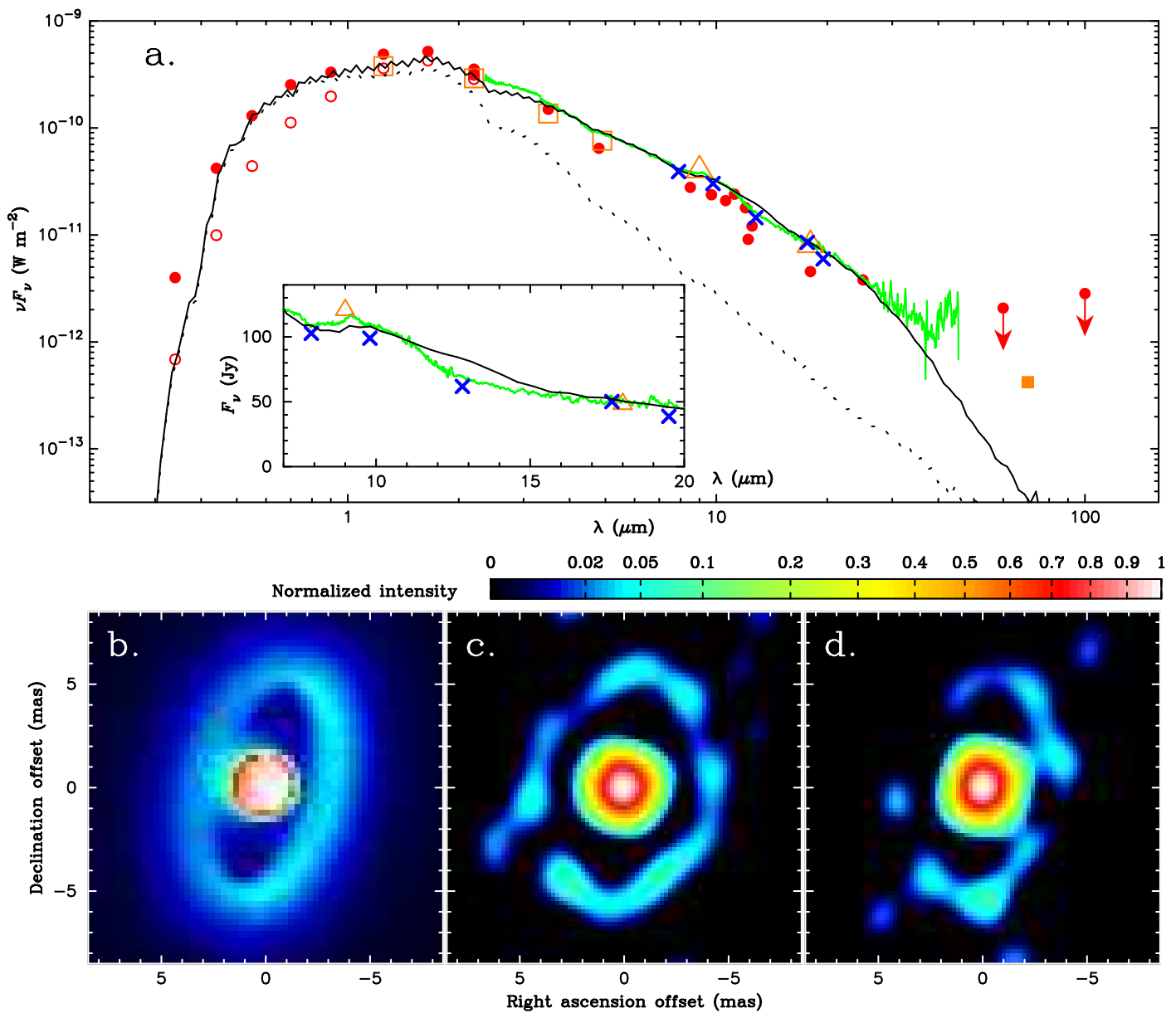}}}
%\resizebox{\hsize}{!}{\rotatebox{0}{\includegraphics{hr3126_model_sed_images_1+3x1.ps}}}
\caption{
Radiative transfer modeling of the observed SED and continuum image of HR3126.
{\bf a:} Comparison of the SED between the best-fit model and the observed
data. 
The solid black line and the dotted line represent the total flux and
stellar flux predicted by the best-fit model, respectively. 
Filled circles show the 
photometric data compiled in Chiar et al. (\cite{chiar93}), which we
dereddened with $A_V = 1.18$ and $R_V = 3.1$ (Sect.~\ref{sect_basic}). 
The original data from Chiar et al. (\cite{chiar93}) 
without dereddening are plotted with the open circles 
(data points longward of 3~\micron\ are not shown 
because the reddening is negligible).
The data taken with COBE/DIRBE, AKARI, and Hershel/PACS are plotted with
the open squares, open triangles, and filled square, respectively. 
The solid green line represents the 
ISO/SWS spectrum. 
The blue crosses show the VISIR measurements of the present work 
(Sect.~\ref{subsect_res_single}).
The inset shows an enlarged view of the mid-infrared wavelength region. 
{\bf b:} 2.2~\micron\ image predicted by the best-fit model. 
{\bf c:} Image reconstructed from the simulated interferometric data
generated from the best-fit model image. 
{\bf d:} Image reconstructed at 2.25~\micron\ in the continuum shown in
Fig.~\ref{hr3126_images}c. The images in panels {\bf b}, {\bf c}, and
{\bf d} are not convolved with a beam. 
}
\label{hr3126_bestmodel}
\end{figure*}

\section{2-D dust radiative transfer modeling}
\label{sect_modeling}

\subsection{Modeling of the SED}
\label{subsect_model_procedure}

We carried out 2-D radiative transfer modeling of the AMBER+GRAVITY 
image in the continuum and the observed SED to derive the physical 
properties of the dust disk, using the Monte-Carlo multi-dimensional 
radiative transfer code \mbox{mcsim\_mpi} (Ohnaka et al. \cite{ohnaka06}; 
\cite{ohnaka16}).
  The absorption of photons and the calculation of the dust temperature
  are treated with the method of Bjorkman \& Wood (\cite{bjorkman01}). 
The scattering of photons by dust grains is incorporated by using the 
scattering matrix (special Mueller matrix). 
The radiation of the central star
is represented by the model of Castelli \& Kurucz (\cite{castelli03}) with 
\TEFF\ = 3750~K, \LOGG\ = 0.5, and [Fe/H] = $-0.5$,
as described in Sect.~\ref{sect_basic}. 
We adopted the disk geometry described as
\[
\rho (r, z) = \rho_0 \left(\frac{r_{\rm in}}{r}\right)^p \exp\left[-\frac{1}{2}\left(\frac{z}{h(z)}\right)^2\right]
\]
\[
h(z) = h_{\rm in} \left(\frac{R}{r_{\rm in}}\right)^{\beta},
\]
where $r$ and $z$ are the radial distance from the star and the height from 
the equatorial plane, respectively, and $R$ is the radial distance in the 
equatorial plane. The parameters $r_{\rm in}$ and $h_{\rm in}$ are the radius 
of the inner rim of the disk and its height at the inner rim, respectively. 
The angular radius of the star of 1.5~mas and the observed semimajor 
axis of 5.3~mas of the inner rim suggest that the inner radius is 
3.5~\RSTAR. 
We tentatively set the disk outer radius to be 500~\RSTAR. 
The free parameters 
to define the physical properties of our disk model 
are the optical depth in the radial 
direction in the equatorial plane ($\tau_V$), 
the height of the disk at its inner rim ($h_{\rm in}$), the radial density 
exponent ($p$), and $\beta$, which characterizes the flaring of the disk. 
In addition, we need two parameters to specify 
the inclination angle ($i$) and the position angle of the disk in the 
plane of the sky (PA). 
It is also necessary to specify the grain properties. 
We adopted the optical properties of the astronomical silicate dust presented
by Ossenkopf et al. (\cite{ossenkopf92}), assuming homogeneous spherical 
grains. The grain radius $a$ is the last free parameter of our modeling. 

We need the bulk density of silicate to calculate the dust mass density
and the total dust mass of the disk.
Bulk densities from 3.0 to 3.7~g~cm$^{-3}$ are adopted in the literature
(e.g., Bouwman et al. \cite{bouwman01}; Suh \cite{suh02}). We adopted
the 3.5~g~cm$^{-3}$ from Draine (\cite{draine03}).

The search for the best-fit model was carried out in the following manner. 
The computation of the model images is more time-consuming than
that of SEDs. Therefore, we first calculated SEDs for a grid of
models with the parameter ranges (\tauV, \hin, $p$, $\beta$, $i$, and $a$) 
listed in Table~\ref{modeltable}.
The disk's position angle in the plane of the sky does not affect the model
SEDs, which is why it was not changed at this step. 
We selected the models that fit the observed SED by visual inspection. 
We did not use the reduced $\chi^2$ for the following reason. 
While only discrete photometric data are available from the visible to the
near-infrared, the ISO/SWS spectrum covers continuously from 2.4 to
45~\micron. Therefore, a model that fits very well to the ISO/SWS spectrum
but fails to reproduce the visible and near-infrared photometric data may still
show a good $\chi^2$ value, which is misleading.

The comparison between the model and observed SEDs shows that 
grains as large as 4~\micron\ can fairly reproduce the broad silicate feature
in the mid-infrared without a prominent peak at $\sim$9.8~\micron\ often seen
in oxygen-rich AGB stars. 
As shown in Fig.~\ref{hr3126_bestmodel}a (inset),
the silicate feature predicted with 4 \micron\ grains is blunt as seen
in the observed spectrum. The model predicts the feature to be too broad,   
failing to reproduce the spectrum between 11.5 and 15~\micron. 
This may be due to the possible presence of smaller grains shielded by
the optically thick 4 \micron\ grains as mentioned below. 
We found that grain sizes in the range of $4 \pm 0.5$~\micron\ can fairly
reproduce the observed blunt silicate feature.
The grain size larger (or smaller) than this range leads to a silicate
feature too featureless (or too prominent) compared to the observed spectrum. 

Another constraint on the grain size
is provided by the dust temperature at the inner rim of 3.5~\RSTAR. 
Our models show that
the inner rim dust temperature reaches 1740~K 
with a grain size of 0.1~\micron\ even for \tauV\ = 1,
and the temperature is even higher for optically thicker cases. 
With a grain size of 0.5~\micron, the dust temperature at the inner rim
is 1650~K for \tauV\ = 1, which may be marginally acceptable, given the
uncertainties in the dust sublimation temperature
(for example, Hillen et al. \cite{hillen14} derived 
an inner rim temperature of $\sim$1600~K for the disk around 
the post-AGB star 89~Her). 
Grains larger than $\sim$0.5~\micron\ can survive at 3.5~\RSTAR\ due to their 
high albedo in the visible and near-infrared.
The grain size of $\sim$4~\micron\ suggested from the mid-infrared spectrum 
is consistent with this grain size range constrained by the inner rim dust
temperature.

For the models selected based on the SED,
we re-computed the models with many more
photon packages to obtain model images of sufficient quality. 
We computed the visibility
amplitude and closure phase at the observed $u\varv$ points from the model
images and compared them with the observed data.
In this comparison, we changed the disk's position angle
in the plane of the sky (PA) as a free parameter for each model 
and selected the best-fit model by eye. 
We opted not to use the reduced $\chi^2$ value for the following reason.
Some models fit very well to the interferometric observables
in a certain range of the spatial frequency (e.g., the visibility amplitude
at spatial frequencies higher than 100~arcsec$^{-1}$), while the fit is poor 
in other regions of the spatial frequency. Nevertheless, if the fit to the
former region is very good, the reduced $\chi^2$ value can be low, despite
the poor fit in other regions. We selected models that fit the
observed visibilities and closure phases as homogeneously as possible across
the observed spatial frequency range.

\subsection{Modeling of the SED and interferometric data}
\label{subsect_model_results}

Figure~\ref{hr3126_bestmodel} shows a comparison of the image and SED of 
the best-fit model with those observed. Comparison of the visibility and 
closure phase is shown in Fig.~\ref{fit_interf_model}. 
From the best-fit model image, 
we also generated simulated interferometric data with noise similar to
the AMBER+GRAVITY measurements 
at the same $u\varv$ points as the observed data. Then we carried out
image reconstruction from the simulated data
with the same set-up as for the real data using both regularizations. 
The model image reconstructed in this fashion is shown in
Fig.~\ref{hr3126_bestmodel}c (only the image reconstructed with
the pixel difference quadratic regularization is shown because two
regularizations resulted in similar images). 
The best-fit model is characterized by 
$\tau_V$ = 3.0, $h_{\rm in}$ = 0.5~\RSTAR, $\beta$ = 2.5, $p$ = 5, and
the grain size of 4~\micron, as summarized in Table~\ref{modeltable}. 

The observed SED and image are reasonably reproduced by the model
from the visible to $\sim$30~\micron. 
The 60~\micron\ and 100~\micron\ fluxes measured with IRAS (data points
with the arrows in Fig.~\ref{hr3126_bestmodel}a) are significantly 
higher than predicted by the model. This is attributed to the
background cirrus not physically associated with HR3126, 
as Chiar et al. (\cite{chiar93}) note. 
However, the model also underestimates the 70~\micron\ Herschel/PACS flux. 
The 70~\micron\ Herschel image in the
Herschel data archive shows that most of the 70~\micron\ flux originates
from the equatorial region and the nebula
extending to a radius of $\sim$1\arcmin\ ($\sim \! 4 \times 10^4$~\RSTAR),
much larger than the outer radius of our models (500~\RSTAR).
Because we focus on the modeling of the inner region of the 
equatorial disk, it is beyond the scope of the present paper to model
the entire disk and nebula, where the disk geometry and grains properties
may be different from the inner disk modeled here.

The value of $\beta$ = 2.5 means that the disk is strongly 
flared, while its density falls off steeply in the radial direction 
as suggested by $p$ = 5.
However, because we adopted the above flared disk density distribution
for simplicity, other disk geometries may be possible. This point
should be further investigated by high spatial resolution thermal-infrared 
imaging with VLTI/MATISSE (Lopez et al. \cite{lopez14}),
which allows us to probe the region farther out than imaged
with AMBER and GRAVITY.

The best-fit inclination angle is derived to be 55\degr\ 
(i.e., the disk is seen at 55\degr\ from pole-on), and the position angle of 
the minor axis of the projected inner rim is 75\degr, roughly aligned
to the axis of the bipolar nebula. 
The dust temperature reaches 1590~K at the inner boundary, which is 
acceptable for a dust sublimation temperature. 
The eastern side of the inner rim is fainter than the western side, as 
seen in the reconstructed image (Fig.~\ref{hr3126_images}). 
This is because the irradiated inner wall is directly seen on the western 
side (far side of the inner rim), while it is hidden from the observer 
on the eastern side (near side of the inner rim). 
The uncertainties in the disk parameters were estimated by changing them
around the best-fit model with steps smaller than the grid in
Table~\ref{modeltable}. 
The estimated uncertainties are 
$\pm 0.5$ ($\tau_V$), $\pm 0.25$ ($h_{\rm in}$), $\pm 0.5$ ($\beta$), 
and $\pm 1$ ($p$). The inclination angle and the position angle of the
disk projected onto the sky are constrained with an uncertainty of
$\pm 5$\degr.
As described above, the uncertainty in the grain size
is $\pm 0.5$~\micron, but as we discuss below, the presence of smaller
grains cannot be excluded.
Because of the steep radial density gradient, 
the uncertainty in the outer radius does not affect the results. 

We also estimated the pressure scale height at the inner rim expected in
hydrostatic equilibrium, which is written as
\[
h_{\rm p} = \sqrt{\frac{kT_{\rm in}r_{\rm in}^3}{\mu_{\rm g} m_{\rm p} G \MSTAR}},
\]
where $T_{\rm in}$, $\mu_{\rm g}$, $m_{\rm p}$, and \MSTAR\ represent
the dust temperature at the inner rim, the mean molecular weight, 
the proton mass, and the current stellar mass,
respectively (e.g., Dullemond et al. \cite{dullemond01}).
The inner rim dust temperature $T_{\rm in}$ and the radius $r_{\rm in}$
were set to 1590~K and 2~au (3.5~\RSTAR), 
respectively, as described above. 
The stellar mass \MSTAR\ is $2.2 \pm 0.7$~\MSOL. 
The mean molecular weight was estimated as follows. 
The dust mass density of the best-fit model at the inner rim in the
equatorial plane is $5.8 \times 10^{-12}$~g~cm$^{-3}$ with the adopted
bulk density of 3.5~g~cm$^{-3}$. Adopting gas-to-dust ratios of 100--200,
the gas density is estimated to be
(5.8--11.6)$\times 10^{-12}$~g~cm$^{-3}$. Then, assuming chemical
equilibrium, we calculated the partial pressure of various molecular and 
atomic species. 
The mean molecular weight is derived to be 2.2 for the above gas density
range. The pressure scale height at the inner rim is estimated to be
$0.27 \pm 0.05$~\RSTAR, which is in fair agreement with the
$0.5 \pm 0.25$~\RSTAR\ derived from the 2-D radiative transfer modeling.

\subsection{Inner rim dust temperature from spectral fitting}
\label{subsect_model_spectral}

We estimated the inner rim dust temperature by fitting
the observed spectrum with the stellar spectrum and the blackbody
radiation from the inner rim, independent of the
radiative transfer modeling. 
As a proxy of the central star of HR3126, 
we used the spectrum of the M2II star HD23475 obtained with the
Infrared Telescope Facility (IRTF) presented by 
Rayner et 
al. (\cite{rayner09})\footnote{http://irtfweb.ifa.hawaii.edu/\~{}spex/IRTF\_Spectral\_Library/index.html} 
because its spectral type and luminosity class are similar to those of
HR3126. 
The spectrum of HD23475 originally taken with a spectral resolution of 2000
was convolved down to match the spectral resolution of 1500 of
the observed spectrum of HR3126 shown in Fig.~\ref{hr3126_spec}.

\begin{figure}
\resizebox{\hsize}{!}{\rotatebox{0}{\includegraphics{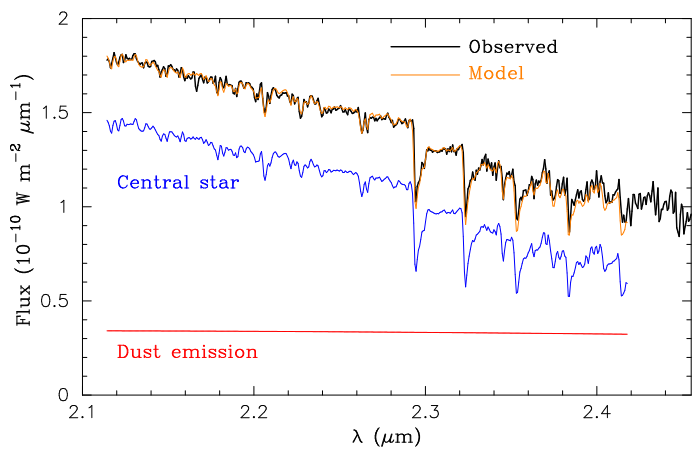}}}
%\resizebox{\hsize}{!}{\rotatebox{0}{\includegraphics{hr3126_specfit.ps}}}
\caption{
  Fitting to the observed spectrum of HR3126 with the central star and
  dust emission components.
  The solid black and orange lines represent the spectrum obtained with AMBER
  and the model spectrum, respectively. 
  The flux of the central star
  and dust at the inner rim (with a dust temperature of 1400~K) 
  predicted by the model
  is shown with the solid blue and red lines, respectively.
  The model spectra are calculated only up to 2.42~\micron, beyond which 
  the IRTF spectrum of the proxy star of HR3126 is not available. 
}
\label{hr3126_specfit}
\end{figure}

The predicted spectrum ($F_{\rm model} (\lambda)$) can be written as the
weighted sum of the flux of the central star ($F_{\star}(\lambda)$,
IRTF spectrum of HD23475) 
and the blackbody radiation from dust 
($B(\lambda,T_{\rm d})$, $T_{\rm d}$ is the dust temperature) 
\[
F_{\rm model}(\lambda) = c_{\rm \star} F_{\star} (\lambda) + c_{\rm d}
B(\lambda, T_{\rm d}), 
\]
\[
c_{\star} = \frac{w_{\star} F_{\rm obs}(\lambda_0)}{F_{\star}(\lambda_0)},
\]
\[
c_{\rm d} = \frac{w_{\rm d} F_{\rm obs}(\lambda_0)}{B(\lambda_0,T_{\rm d})},
\]
where $\lambda_0$ is a reference wavelength at which the flux of the
model spectrum is matched to the observed flux, and $w_{\star}$ and
$w_{\rm d}$ represent the fraction of the stellar flux and dust emission
at $\lambda_0$, respectively ($w_{\star} + w_{\rm d} = 1$). 
We set $\lambda_0$ to 2.25~\micron\ and derived $w_{\star}$
to be 0.78 from the reconstructed image in the 2.25~\micron\ continuum. 
The dust temperature is the only parameter, and it 
changes not only the slope of the continuum
emission but also the depth of the CO bands of the summed spectrum. 
By changing $T_{\rm d}$ from 1000 to 2000~K, 
we found that the observed spectrum is fitted with
$T_{\rm d} = 1400 \pm 200$~K (Fig.\ref{hr3126_specfit}). This temperature
range is in broad agreement with the inner rim dust temperature of
1590~K derived from the above 2-D radiative transfer modeling.

\begin{table*}
\begin{center}
\caption {
2-D radiative transfer modeling of the dust disk around HR3126.
}
\begin{tabular}{l c c}\hline
Parameter & Range & Best-fit value  \\
\hline
$r_{\rm in}$ : inner radius (\RSTAR) & fixed & 3.5 (= 2~au) \\
$r_{\rm out}$: outer radius  (\RSTAR) & fixed & 500 (= 285~au) \\
\tauV\ (optical depth at 0.55~\micron) & 0.5, 1.0, 2.0, 3.0, ... 7.0 & $3.0 \pm 0.5$ \\
\hin\ : disk height at $r_{\rm in}$ (\RSTAR) & 0.125, 0.25, 0.5, 0.75, ... , 1.5
& $0.5 \pm 0.25$  \\
                                           &  & (= $0.29\pm 0.14$~au)\\
$\beta$: flaring parameter & 0.5, 1.0, 1.5, ..., 3.5     & $2.5 \pm 0.5$   \\
$p$: radial density exponent & 2, 3, 4, ..., 7    & $5 \pm 1$       \\
$a$: grain radius (\micron) & 0.1, 0.5, 1.0, 3.0, 4.0, 5.0, 6.0
& $4.0 \pm 0.5$ \\
$i$: inclination (\degr) & 30, 35, 40, ... , 80& $55 \pm 5$\\
PA: position angle (\degr) & 60, 65, 70, ... , 100 & $75 \pm 5$\\
\multicolumn{3}{l}{(Symmetry axis of the disk projected onto the plane of the sky)}\\
Disk dust mass (\MSOL) & --- & $\ge 2.6 \times 10^{-6}$ \\
Disk total mass (\MSOL) & --- & $\ge 5.2 \times 10^{-4}$ \\
\multicolumn{3}{l}{(gas-to-dust ratio = 200)}\\
\hline
\label{modeltable}
\end{tabular}
\end{center}
\end{table*}

\section{Discussion}
\label{sect_discuss}

\subsection{Properties of the equatorial disk}
\label{subsect_disk}

Our radiative transfer modeling shows the presence of grains as large as 
4~\micron\ at only 3.5~\RSTAR\ from the central star. 
To see whether the grains can exist in a stable manner in the disk, 
we estimate the ratio of the radiation pressure to the gravitational acceleration 
$\beta_{\rm dust} \equiv {a_{\rm rad}}/{\varg_{\rm grav}}$, where 
$a_{\rm rad}$ and $\varg_{\rm grav}$ denote the acceleration due to the radiation 
pressure and the gravitational acceleration, respectively. 
The ratio $\beta_{\rm dust}$ is given by
\[
\beta_{\rm dust} = 1146 \frac{L_{\star}}{10^4 \LSOL} \frac{1}{M_{\star} (\MSOL)} 
\frac{Q}{0.2} \frac{1}{a (\micron)} \frac{1}{\rho_{\rm bulk} ({\rm g \, cm^{-3})}}, 
\]
where $L_{\star}$, $M_{\star}$, $Q$, $a$, and $\rho_{\rm bulk}$ are 
the luminosity of the central star, its mass, the flux-mean opacity, 
the grain radius, and the bulk density of dust grains, respectively 
(Yamamura et al. \cite{yamamura00}). 
We obtained $Q \approx 2.2$ from the extinction coefficients of 
the 4~\micron\ grain and the spectrum of the central star used in our 
radiative transfer modeling. 
Inserting $L_{\star}$ = 2500~\LSOL, $M_{\star}$ = 2.2~\MSOL,
$a$ = 4.0~\micron, 
and $\rho_{\rm bulk}$ = 3.5~g~cm$^{-3}$ results in $\beta_{\rm dust}$ = 102. 
The value of $\beta$ for the gas-dust-mixture is diluted by the 
gas-to-dust ratio $f$, that is, $\beta = \beta_{\rm dust} / f$. 
While a gas-to-dust ratio of 100 is often assumed in the literature,
higher gas-to-dust ratios of 200--500 are also found or adopted in
Galactic sources (e.g., Ramstedt et al. \cite{ramstedt08};
Olofsson et al. \cite{olofsson19}).  
With $f$ = 100--200 adopted in the present work,
$\beta$ is estimated to be 0.5--1.0. 
Therefore, as far as the bulk density of $\ga$3.5~g~cm$^{-3}$ is assumed, 
the large 4-\micron\ grains can exist in the dust disk 
in a stable manner, not blown away 
by the radiation pressure from the central star. 
If the bulk density of 3~g~cm$^{-3}$ and the low gas-to-dust ratio of 100 
are adopted, $\beta$ is
1.2, which indicates that the disk is expanding. 
In this case, the disk material should be replenished by the stellar wind
from the central M giant. However, its current mass-loss rate 
is low, $(2-4) \times 10^{-7}$~\MSOLPERYR\ (Sect.~\ref{sect_basic}),
which implies an optically thin stellar wind (e.g., Heras \& Hony
\cite{heras05}). Given the difficulty in sustaining 
the optically thick disk with the optically thin stellar wind, the expanding
disk scenario is not favorable. However, we need high spatial resolution
kinematic study of the disk to confirm this argument.

We found that $\beta$ exceeds 1 for a grain radius of 1~\micron\ 
even with the rather high gas-to-dust ratio of 500. 
Therefore, grains smaller than $\sim$1~\micron\ should be blown away 
by the radiation pressure and therefore should be absent in the disk.
However, it is possible that the small grains might 
exist in the disk if they are shielded from the central star's radiation 
by the optically thick inner rim. 
High spatial resolution data in the mid-infrared would be useful to 
prove or disprove this possibility.

Recent visible and near-infrared polarimetric observations show that 
grains that form in the stellar winds of oxygen-rich AGB stars are 
0.1--0.4~\micron\ (Norris et al. \cite{norris12}; Khouri et al. 
\cite{khouri16}, \cite{khouri20}; Ohnaka et al. \cite{ohnaka16}, 
\cite{ohnaka17}; Adam \& Ohnaka \cite{adam19}). 
If the disk formed together with the bipolar nebula due to binary 
interaction, the grains must have grown from 
a few $\times \, 10^{-1}$~\micron\ to 4~\micron\ on a time scale of 
the nebula's dynamical age of $\sim$3900~yrs. 
The disk's dust mass is estimated to be $2.6 \times 10^{-6}$~\MSOL, 
which translates into a total disk mass of 
(2.6--5.2)$\times 10^{-4}$~\MSOL\ with the gas-to-dust ratios of 100--200. 
The total disk mass is three orders of magnitude smaller than
the nebula's mass of $\sim$1~\MSOL\ (L.-\AA. Nyman, priv. comm.).
However, as mentioned above, our model underestimates the 70~\micron\
Herschel/PACS flux,
suggesting the presence of cold dust in the outer region of the disk.
Therefore, the above disk mass is a lower limit, and at the moment
we cannot draw a definitive conclusion about the fraction of mass
that formed the disk.

\subsection{Possibility of a companion}
\label{subsect_companion}

As mentioned in Sect.~\ref{subsect_res_vlti}, the presence of a companion 
is neither confirmed nor disproved for HR3126 in the literature. 
It is possible that a companion was shredded into 
an accretion disk around the primary star as Nordhaus \& Blackmann 
(\cite{nordhaus06}) postulate as one of the common-envelope evolution 
scenarios. The bipolar nebula might have formed as a consequence of the 
violent mass ejection associated with the disruption of the companion. 
On the other hand, if the companion was not shredded, 
and the envelope (of the primary star) was not entirely 
ejected, it is possible that the matter falls back and form a circumbinary disk 
(e.g., Ivanova et al. \cite{ivanova13}). 
In this case, the companion is likely to exist within the dust cavity.

We attempted to set a constraint on the separation between the red
giant and a putative companion in the following manner. 
Artymowicz \& Lubow (\cite{artymowicz94}) show that the ratio of
the tidally truncated inner radius of the circumbinary disk
to the binary semimajor axis is $\sim$1.7 
for mass ratios between 0.05 and 0.5 in the case of circular orbits.
The ratio can be larger in the case of eccentric orbits. 
The ratio of 1.7 translates into a binary separation of 2.1~\RSTAR\ (1.2~au) 
for the radius of HR3126's inner rim of 3.5~\RSTAR\ (2~au). 
The binary separation cannot be greater than 2.1~\RSTAR\ because
the tidal truncation radius would be larger than the observed inner rim
radius.   
The binary separation can be smaller than 2.1~\RSTAR, in which case
the inner rim radius is determined by dust sublimation. 
The binary separation of $\la \! 2.1$~\RSTAR\ (1.2~au) is very small,
given that
non-Mira-type K and M giants possess a dense molecular atmosphere extending
out to $\sim$2~\RSTAR\ (e.g., Ohnaka \cite{ohnaka13}; Ohnaka \& Morales
Mar\'in \cite{ohnaka18}; Ohnaka et al. \cite{ohnaka19}).
A binary companion orbiting through the dense outer atmosphere of
the red giant would
experience considerable friction, which might make the companion
spiral in to the primary red giant.
Therefore, the presence of a companion is disfavored,
although we cannot yet entirely exclude its presence. 

If a companion is nevertheless present,
it is likely to be a main-sequence star less massive than the red giant. 
If we assume that the red giant's initial mass is 3~\MSOL\
(Sect.~\ref{sect_basic}),
and the main-sequence companion is less massive than 2~\MSOL, the luminosity
of the companion should be lower than 20~\LSOL\ based on 
the theoretical evolutionary tracks
of Lagarde et al. (\cite{lagarde12}).
This means that 
the companion's luminosity is more than 100 times lower than that of the
red giant.
The dynamic range of the reconstructed images is estimated to be
$\sim$30 from the reconstruction noise. 
The non-detection of a companion in the reconstructed images implies
that its \mbox{2.2 \micron} brightness is more than $\sim$30 times lower
than that of the red giant. 
Therefore, it is not surprising that a putative companion more than 100
times less luminous than the red giant was not detected 
in the AMBER+GRAVITY images.

If we assume the aforementioned binary separation of 1.2~au, 
the orbital velocity of the red giant and the period are
estimated to be 8~\KMS\ (or 4~\KMS) and 0.9~yr (or 0.9~yr) 
for a 0.5~\MSOL\ (or 0.2~\MSOL) companion.
While these values seem to show
that radial velocity monitoring is feasible to set
constraints on the mass and separation of the companion, 
a problem is wavelength shifts of spectral lines caused by pulsation
and/or convection of the red giant. 
Lebzelter et al. (\cite{lebzelter02}) show that the velocity variation
amplitude in semiregular and irregular variable M giants is 3--4~\KMS.
Therefore, a companion less massive than $\sim$0.2~\MSOL\ would be
difficult to confirm by radial velocity monitoring.

\subsection{Comparison with other red giants associated with optical bipolar nebulae}
\label{subsect_comparison}

As mentioned in Sect.~\ref{sect_intro}, the number of red giants associated
with bipolar nebulae is very limited. 
Particularly, non-interacting systems (interacting systems include,
for example, symbiotic stars) with red giants and optical 
bipolar nebulae are rare.
Frosty Leo and OH231.8+4.2 are well-studied among such objects. 
Both objects have red giant stars at their center and 
exhibit prominent optical bipolar nebulae whose spatial
extent is comparable to that of the Toby Jug Nebula ($\sim$0.4~pc,
from the tip of the one lobe to the tip of the other). 
The central star of Frosty Leo is a red giant with an effective
temperature of $\sim$3750~K (Robinson et al. \cite{robinson92}), and
the bipolar nebula has an angular extent of $\sim$0\farcm4
(Sahai et al. \cite{sahai00}), which corresponds to $\sim$0.4~pc
at a distance of 3~kpc (Mauron et al. \cite{mauron89};
Robinson et al. \cite{robinson92}). 
The central star of OH231.8+4.2 is an M9 Mira star (Cohen \cite{cohen81};
Kastner et al. \cite{kastner92}, \cite{kastner98}), 
and its bipolar nebula has an extension of $\sim$0.4~pc along its
symmetry axis (Bujarrabal et al. \cite{bujarrabal02}). 

However, the inner rim of the equatorial dust disk of both objects
is much larger than that of HR3126. Murakawa et al. (\cite{murakawa08})
derived the inner radius to be 5200~\RSTAR\ ($\approx$1000~au) for
Frosty Leo. Matsuura et al. (\cite{matsuura06}) estimate the inner radius of
OH231+4.2 to be 40--50~au, which translate into 19--24~\RSTAR\ with
\RSTAR\ = 2.1~au (S\'anchez Contreras et al. \cite{sanchez_contreras18}). 
A possible explanation is the expansion of the disks. 
In fact, (sub)millimeter molecular line observations show the
presence of slowly expanding disks in Frosty Leo and OH231.8+4.2
(Castro-Carrizo et al. \cite{castro-carrizo05}; 
S\'anchez \mbox{Contreras} et al. \cite{sanchez_contreras18}). 
However, it is also possible that the disk inner rim radius is
determined by the past binary interaction.

An optical bipolar nebula has recently been imaged toward the 
second closest AGB star \lpup\ (Kervella et al. \cite{kervella15}).
The central red giant of
\lpup\ is very similar to HR3126 in terms of the effective temperature
and the linear radius. However, the optical bipolar nebula of \lpup\ is
seen only up to $\sim$15~au, much smaller than the 0.2~pc of HR3126.
The dust disk of \lpup\ has an inner radius of 6~au
(10~\RSTAR\ using the stellar radius of 
123~\RSOL\ derived by Kervella et al. \cite{kervella14}), significantly
larger than the 2~au (3.5~\RSTAR) of HR3126.
The grain size also significantly differs
between two objects: 0.1--0.3~\micron\ in \lpup\
(Kervella et al. \cite{kervella14}; \cite{kervella15})
and $\sim$4~\micron\ in HR3126.

Binary interaction is considered to be the key in the formation of the
bipolar outflow and disks in these four objects (HR3126, OH231.8+4.2,
Frosty Leo, and \lpup). In fact, a binary companion 
is detected in
OH213.8+4.2 and \lpup\ (S\'anchez Contreras et al. \cite{sanchez_contreras04};
Kervella et al. \cite{kervella15}), while no companion has been confirmed in
HR3126 and Frosty Leo. 
It is very likely that 
the differences in the properties of the bipolar nebula and equatorial disk
among four objects result 
from the differences in the properties of their companion and the history
of the interaction. While it is not possible to disentangle the binary
interaction history of HR3126 from the present observations alone,
further high-resolution observations of this object will be useful
for better understanding the role of binary interaction in the formation
of the bipolar outflow and the equatorial disk. It should be noted that
while the central star of OH231.8+4.2, Frosty Leo, and \lpup\ is
partially or completely obscured by the equatorial dust disk, 
HR3126 allows us to directly observe the central region close to the star. 
Therefore, further high spatial resolution observations of this target
will provide information about the formation mechanism of the bipolar
nebula and the equatorial disk, complementary to the above three objects.

\section{Conclusion}
\label{sect_concl}

Using the VLTI instruments AMBER and GRAVITY, 
we imaged the central region of the Toby Jug Nebula (reflection nebula 
IC2220), which is an arcminute-scale bipolar outflow emanating
from the AGB star HR3126.  The images reconstructed in the continuum 
at $\sim$2.2~\micron\ reveal the inner rim of the equatorial dust disk, 
whose axis is approximately aligned with the axis of the
optical bipolar nebula. 
The images reconstructed in the CO first overtone bands show that the 
emission from inside the dust cavity is elongated in the N-S direction. 
This indicates the oblateness of the central star's atmosphere in the 
equatorial plane or the presence of CO gas in the disk inside the dust 
cavity. 

Our single-dish VLT observations with SPHERE-ZIMPOL, 
NACO, and VISIR show that HR3126 is unresolved at 0.55, 2.24, 
and 7.9--19.5~\micron\ at the resolutions of 25~mas, 64~mas, and
0\farcs4--0\farcs6, respectively. These results suggest the absence of
an extended component larger than the field of view of AMBER and GRAVITY
of 250~mas. 

The observed SED and interferometric image of the resolved disk
in the continuum can be 
explained by an optically thick flared disk model with an inner rim radius 
of 3.5~\RSTAR\ (= 2~au), consisting of silicate grains as large as 4~\micron. 
Such large grains are likely to exist in a stable manner,
not blown away by radiation pressure. 
Also, the derived grain size is much larger than the 0.1--0.4~\micron\ found 
in cool oxygen-rich AGB stars, suggesting efficient grain growth 
on a time scale of the nebula's dynamical age of $\sim$3900~yrs. 
The total disk mass estimated from our modeling is 
$\sim \! 5\times 10^{-4}$~\MSOL\ with a gas-to-dust ratio of 200.
However, this is a lower limit, given the presence of 
cold dust in the outer region of the disk, not included in our modeling. 
High spatial resolution observations in the thermal infrared and
in the (sub)mm domain with VLTI/MATISSE and ALMA 
are valuable for probing the region farther
out than imaged with AMBER and GRAVITY. 

The non-detection of a companion in the reconstructed images suggests that
its 2.2~\micron\ brightness is more than $\sim$30 times lower than that of
the red giant or it might have been shredded due to binary interaction
responsible for the formation of the bipolar nebula and the disk.

\begin{acknowledgement}
  We thank the ESO Paranal team for supporting our VLT and VLTI observations.
  We are also grateful to Lars-\AA ke Nyman for fruitful discussions about
  the modeling of their radio data.  
K.O. acknowledges the support of the Agencia Nacional de 
Investigaci\'on y Desarrollo (ANID) through
the FONDECYT Regular grant 1180066. 
This research made use of the \mbox{SIMBAD} database, 
operated at the CDS, Strasbourg, France. 
This publication makes use of data products from the Two Micron All Sky
Survey, which is a joint project of the University of Massachusetts and the
Infrared Processing and Analysis Center/California Institute of Technology,
funded by the National Aeronautics and Space Administration and the National
Science Foundation.
\end{acknowledgement}

\appendix

\section{VLTI/AMBER and GRAVITY observations}

\subsection{Comparison of the AMBER and GRAVITY data}
\label{appendix_amber_gravity}

\begin{figure}
\resizebox{\hsize}{!}{\rotatebox{0}{\includegraphics{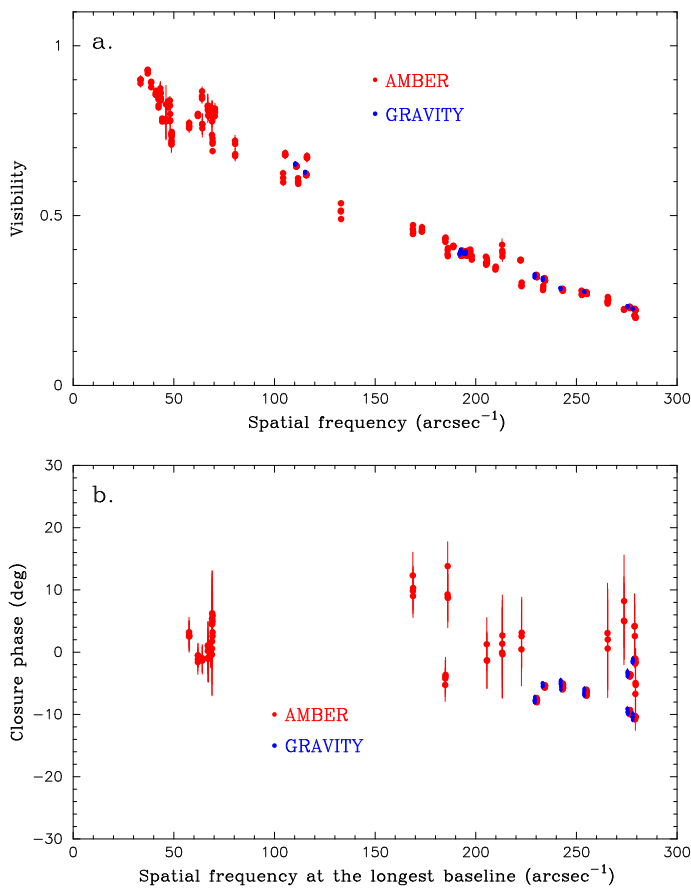}}}
%\resizebox{\hsize}{!}{\rotatebox{0}{\includegraphics{hr3126_vis_amber_gravity.ps}}}
\caption{
  Comparison of the AMBER and GRAVITY data of HR3126.
  {\bf a:} Visibility. {\bf b:} Closure phase.
  In each panel the red and blue dots represent the data measured with
  AMBER and GRAVITY, respectively.  
}
\label{hr3126_vis_amber_gravity}
\end{figure}

Figure~\ref{hr3126_vis_amber_gravity} shows a comparison of the
visibilities and closure phases measured with AMBER and GRAVITY.
The consistency of the AMBER and GRAVITY data means no significant
time variation and the reliability of the data calibration. 

%\clearpage
\subsection{Spectroscopic calibration of the AMBER and GRAVITY spectra}
\label{appendix_spec}

We carried out 
the spectroscopic calibration to remove the telluric lines and instrumental
effects as 
$F_{\rm sci}^{\rm true} = F_{\rm sci}^{\rm obs} \times F_{\rm cal}^{\rm true}/F_{\rm cal}^{\rm obs}$, 
where $F_{\rm sci (cal)}^{\rm true}$ and $F_{\rm sci (cal)}^{\rm obs}$
denote the true and observed spectra of the science target (HR3126) or 
the calibrator ($\beta$~Pyx or $\iota$~Car), respectively. 
Because the true spectrum of $\beta$~Pyx or $\iota$~Car is not available 
in the literature, we used the high-resolution 
($\lambda/\Delta \lambda = 45000$) spectrum of HD219477 and HD6130 
obtained by Park et al. (\cite{park18})\footnote{http://starformation.khu.ac.kr/IGRINS\_spectral\_library} as their proxies because their 
spectral types of G5II/III (HD219477) and F0II (HD6130) are the same 
or close to those of $\beta$~Pyx and $\iota$~Car. The spectra of these 
proxy stars were convolved to match the spectral resolution of 1500 
and used for the spectroscopic calibration described above. 
The absolute flux calibration (i.e., photometric calibration) of the spectra 
of HR3126 derived from the AMBER and GRAVITY data was done using the 
ISO/SWS spectrum (Sloan et al. \cite{sloan03}). 
The AMBER and GRAVITY spectra were scaled to match the ISO/SWS spectrum 
in the overlapping wavelength region between 2.36 and 2.45~\micron.

\section{VLT single-dish observations}
\label{appendix_obs_single}

\subsection{Visible polarimetric imaging with SPHERE-ZIMPOL}
\label{appendix_obs_zimpol}

VLT/SPHERE is an instrument for high spatial resolution and high contrast
imaging from 0.55 to 2.32~\micron,
equipped with an extreme adaptive optics (AO) 
system (Beuzit et al. \cite{beuzit08}). 
We used the ZIMPOL module (Thalmann et al. \cite{thalmann08}) 
for nearly diffraction-limited visible polarimetric imaging of the central 
region of HR3126. 
Our SPHERE-ZIMPOL observations (Program ID: 096.D-0482) 
took place on 2016 March 15 (UTC) in P2 mode, which keeps the field 
orientation fixed.  
The SPHERE-ZIMPOL instrument allows us to use two different or same 
filters for its two cameras. We use the filter V (central wavelength = 
554~nm, FWHM = 80.6~nm) for both cameras.
The field of view was 2\arcsec $\times$ 2\arcsec\ with a pixel scale of
3.628~mas. 
The K3/4III star HD 65273 ($V$ = 5.6) was observed as a 
point spread function (PSF) reference star. 
Given its estimated $V$-band angular diameter of 1.35~mas (Bourges et al. 
\cite{bourges17}), it is entirely unresolved with the spatial 
resolution of SPHERE-ZIMPOL. 
For each of HR3126 and the PSF reference star, we took $N_{\rm exp}$ exposures 
for each of the Stokes $Q_{+}$, $Q_{-}$, $U_{+}$, and $U_{-}$ components, 
with NDIT frames contained in each exposure. 
The summary of our SPHERE-ZIMPOL observations is given in 
Table~\ref{obs_sphere_log}.
HR3126 was also observed on another three nights, 
2015 November 20, 2015 November 27, and 2016 January 30. However, the 
PSF reference data taken on 2015 November 20 show pronounced instrumental 
artifacts. The data taken on the remaining two nights were classified as 
``C'' by ESO, which means that the data quality is not guaranteed. Therefore, 
we discarded these data.

The data were reduced with the SPHERE pipeline 
ver.~0.38.0-9\footnote{http://www.eso.org/sci/software/pipelines/sphere} 
in the same manner as in our previous works (Ohnaka et al. 
\cite{ohnaka16}, \cite{ohnaka17}; Adam \& Ohnaka \cite{adam19}). 
The data taken with two cameras of the ZIMPOL module were averaged. 
The V-filter Strehl ratios both for 
HR3126 and the PSF reference star, which are listed in 
Table~\ref{obs_sphere_log}, were estimated from the $H$-band Strehl ratios 
recorded in the GENSPARTA FITS files using the Mar\'echal 
approximation as described in Adam \& Ohnaka (\cite{adam19}). 
The V-filter Strehl ratios for HR3126 and the PSF reference star are 
0.39 and 0.31, respectively.

\subsection{Near-infrared speckle interferometry with NACO}
\label{appendix_obs_naco}

We carried out speckle interferometric observations of HR3126 with 
the VLT/NACO instrument. 
We used the fast read-out of the instrument in no-AO mode (i.e., 
adaptive optics was turned off), which allowed us to extract interferometric 
data at baselines shorter than 8~m from a large number of short exposures. 

Our NACO speckle observations of HR3126 (Program ID: 099.D-0493) 
occurred on 2017 April 10--11, using 
the IB2.24 filter centered at 2.24~\micron\ with a
FWHM of 0.06~\micron.  
We used the S27 camera with a pixel scale of 27 mas, 
applying the hardware windowing with 512$\times$514 pixels with 
DIT = 150~ms. 
As a calibrator, we observed $\beta$~Vol (K2III, angular diameter = 2.88~mas, 
Bourges et al. \cite{bourges17}), which appears to be a point source 
with the spatial resolution of NACO at 2.24~\micron. 
A summary of our NACO observations is given in Table~\ref{obs_naco_log}. 

We reduced the NACO data with the bispectrum speckle interferometry method 
to extract the 2-D visibility and the bispectrum 
(Weigelt \cite{weigelt77}; Lohmann et al. \cite{lohmann83}; 
Hofmann \& Weigelt \cite{hofmann86}).
While the entire field of view on the detector was 
14\arcsec$\times$14\arcsec, the interferometric observables were extracted from 
a field of view of 3.5\arcsec$\times$3.5\arcsec\ centered around the target. 
We discarded the data at baselines between 7 and 8~m 
(telescope's diameter) because of the poor quality. 
The spatial resolution corresponding to the maximum baseline of 7~m is
64~mas.

\subsection{Mid-infrared imaging with VISIR}
\label{appendix_obs_visir}

Using the VISIR instrument (Lagage et al. \cite{lagage04}), 
we obtained mid-infrared imaging data of HR3126 in the $N$ and $Q$ bands 
(Program ID: 096.D-0482). 
As summarized in Table~\ref{obs_visir_log}, HR3126 was observed with three 
filters in the $N$ band and two filters in the $Q$ band: 
J7.9 (central wavelength $\lambda_c$ = 7.72~\micron, 
FWHM = 0.56~\micron), SIV\_1 ($\lambda_c$ = 9.81~\micron, FWHM = 
0.19~\micron), NeII ($\lambda_c$ = 12.8~\micron, FWHM = 0.2~\micron), 
Q1 ($\lambda_c$ = 17.65~\micron, FWHM = 0.83~\micron), and 
Q3 ($\lambda_c$ = 19.50~\micron, FWHM = 0.40~\micron). 
We observed Canopus ($\alpha$~Car, A9II) with the same filters as a 
PSF reference star and also for flux calibration. 
The pixel scale and field of view of our VISIR data are 45~mas and 
38\arcsec $\times$ 38\arcsec, respectively. 
The observations were carried out with chopping and nodding to subtract 
the sky background, with the chopping and nodding angles both set to 
7\arcsec. 

We removed the sky background by subtracting the chopped and nodded images. 
The resulting sky-subtracted images were then recentered and co-added. 
From these recentered and co-added images, we further removed 
the residual of the sky background not entirely removed in the first step 
by subtracting the average pixel values in the areas sufficiently far away 
from the objects. 

For the flux calibration of the HR3126 data, we first computed 
the flux of the calibrator Canopus with each filter, using its 
ISO/SWS spectrum (Sloan et al. \cite{sloan03}) and the transmission 
function of each filter\footnote{Available at https://www.eso.org/sci/facilities/paranal/instruments/\\visir/inst.html}. 
With the images of Canopus flux-calibrated with the resulting fluxes, we 
measured the flux of HR3126 with each filter within an aperture radius of 
1\farcs5. 
The uncertainty in the flux was estimated by computing the fluxes from 
the data (both science and calibrator data) split into each nodding cycle. 
The measured fluxes of HR3126 are listed in Table~\ref{res_visir_table}.

\section{Image reconstruction from AMBER and GRAVITY data}
\label{appendix_reconst}

\begin{figure}
\resizebox{\hsize}{!}{\rotatebox{0}{\includegraphics{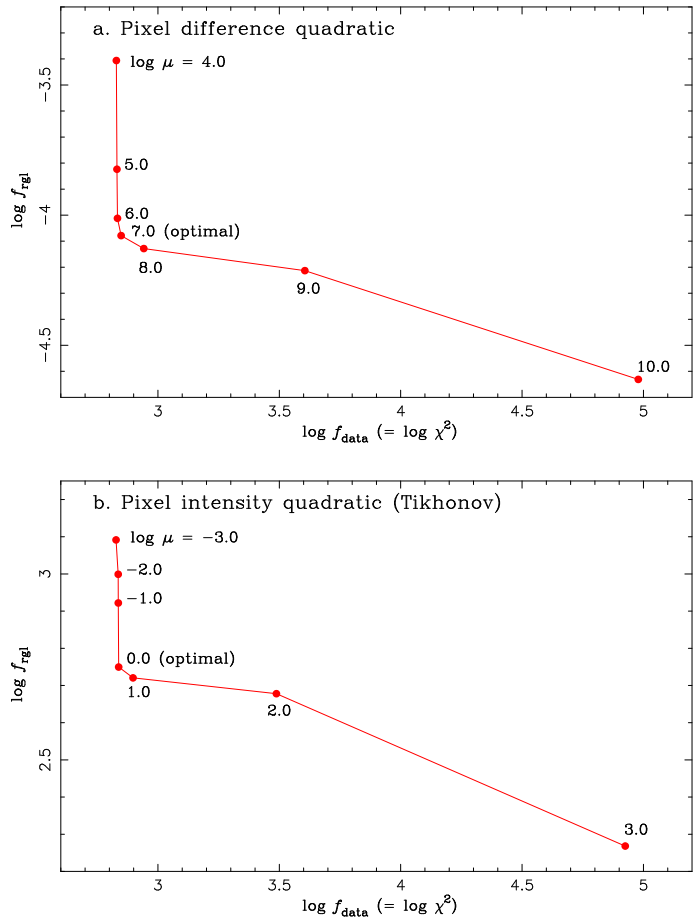}}}
%\resizebox{\hsize}{!}{\rotatebox{0}{\includegraphics{l-curve_1x2.ps}}}
\caption{
L-curves derived with the pixel difference quadratic (panel {\bf a})
and pixel intensity quadratic (panel {\bf b}) regularizations
at 2.25~\micron\ in the continuum. 
$f_{\rm rgl}$ and $f_{\rm data}$ represent the regularization term
and data term (this latter is equal to $\chi^2$), respectively. 
The number beside each dot is the logarithm of the hyperparameter
$\mu$. 
}
\label{l-curves}
\end{figure}

The L-curves derived with the pixel difference quadratic and
pixel intensity quadratic regularizations are shown in
Fig.~\ref{l-curves}. 

\begin{figure*}
\resizebox{\hsize}{!}{\rotatebox{-90}{\includegraphics{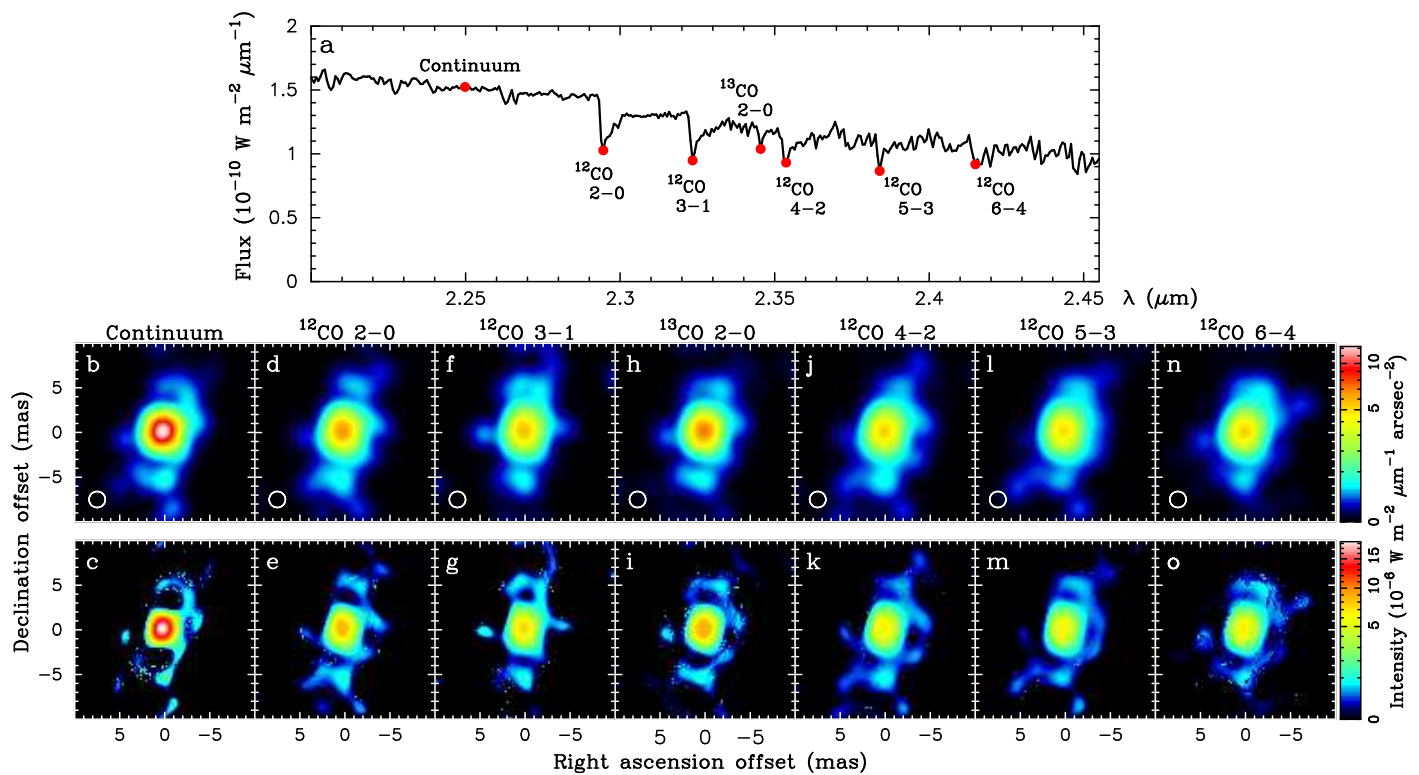}}}
%\resizebox{\hsize}{!}{\rotatebox{-90}{\includegraphics{hr3126_images_quadratic_7x2_sqrtcolorbar.ps}}}
\caption{
Wavelength-dependent images of the central region of the Toby Jug Nebula
around the AGB star HR3126 
reconstructed with the pixel intensity quadratic (Tikhonov) regularization, 
shown in the same manner as in Fig.~\ref{hr3126_images}.
}
\label{hr3126_images_quadratic}
\end{figure*}

We show in Fig.~\ref{hr3126_images_quadratic} 
the images reconstructed with the pixel intensity quadratic regularization.

\begin{figure*}
\begin{center}
\resizebox{13.cm}{!}{\rotatebox{0}{\includegraphics{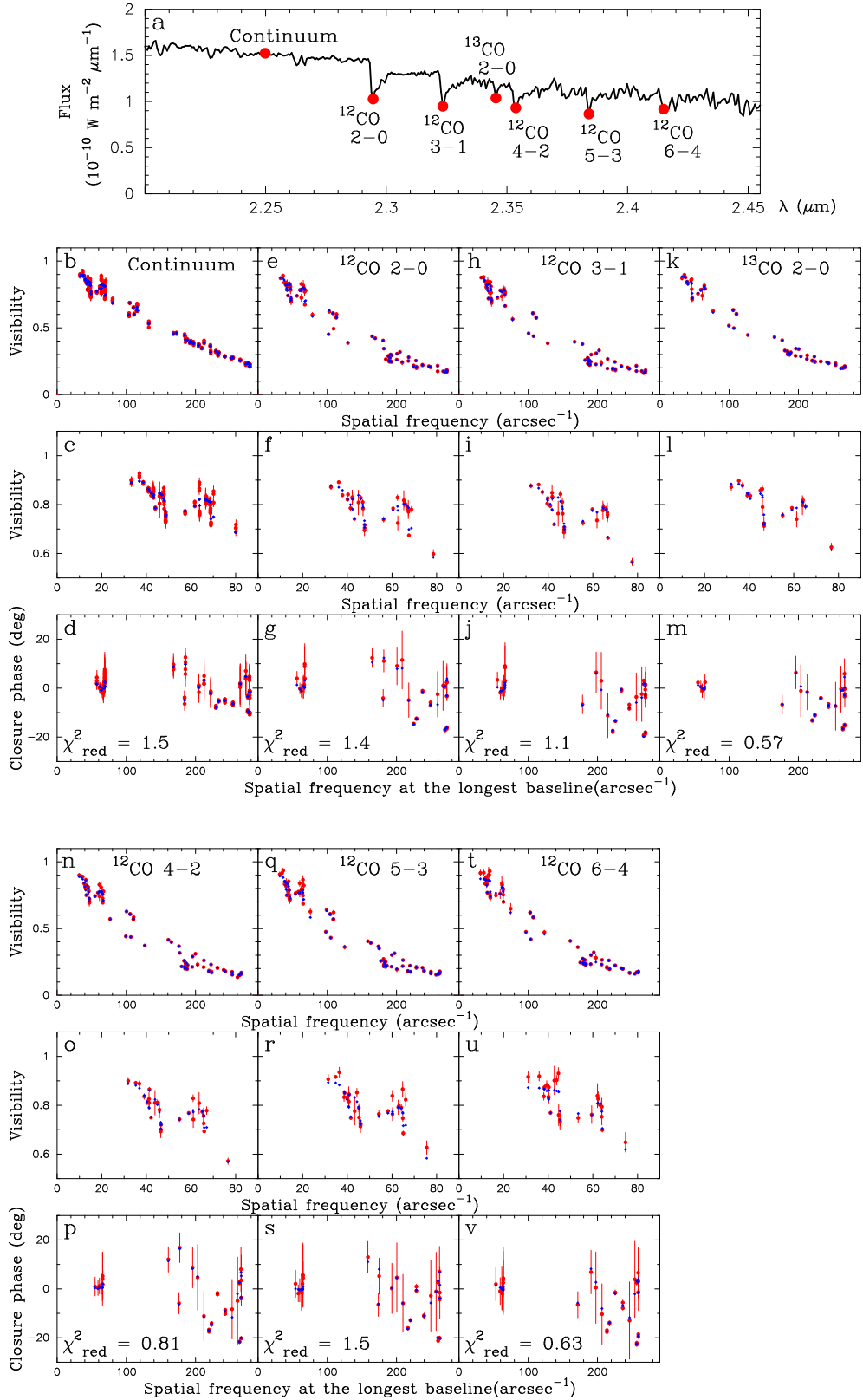}}}
%\resizebox{13.cm}{!}{\rotatebox{0}{\includegraphics{fit_interf_1+3x9.ps}}}
\end{center}
\caption{
Comparison of the observed interferometric observables with those computed 
from the (unconvolved) images of HR3126 reconstructed with the pixel
difference quadratic regularization. 
{\bf a:} Observed spectrum of HR3126. The filled dots represent the 
wavelength channels at which the comparison of the interferometric observables 
is shown in the panels below.
{\bf b}--{\bf m:} Columns each show the comparison of the visibility, 
visibility at low spatial frequencies, and closure phase in the continuum 
and three CO band heads ($^{12}$CO 2 -- 1, 3 -- 1, and $^{13}$CO 2 -- 0). 
The measurements and the values computed from the reconstructed images 
are plotted with the red symbols with the error bars and the blue symbols, 
respectively. 
The total reduced $\chi^2$ value of the fit (including both the visibility
and closure phase) is also shown in each column. 
{\bf n}--{\bf v:} Comparison in the $^{12}$CO 4 -- 2, 5 -- 3, and 6 -- 4 
band heads shown in the same manner as in panels {\bf b}--{\bf m}.
}
\label{fit_interf}
\end{figure*}

Figure~\ref{fit_interf} shows a comparison of the measured interferometric
observables (visibility and closure phase) and those from the images
reconstructed with the pixel difference quadratic regularization.

\section{Results of VLT single-dish observations}
\label{appendix_res_single}

\subsection{SPHERE-ZIMPOL visible polarimetric imaging}
\label{appendix_res_zimpol}

We derived the FWHMs of the azimuthally averaged radial 
intensity profiles of HR3126 and the PSF reference star HD 65273 to see
if HR3126 is more extended than the PSF reference. 
As listed in Table~\ref{obs_sphere_log}, 
the image of the PSF reference star is slightly broader than that of 
HR3126 owing to the slightly lower AO performance on the PSF reference star.
This can be seen in the lower Strehl ratio of the PSF reference data 
compared to the HR3126 data. 
Therefore, HR3126 is unresolved at 0.55~\micron\ 
with the spatial resolution of 25~mas of our ZIMPOL observations 
(FWHM of the PSF reference star's image). 
We found no significant polarimetric signals in the polarized 
intensity map and the polarization degree map either.

\subsection{NACO 2.24~\micron\ speckle interferometry}
\label{appendix_res_naco}

\begin{figure}
\begin{center}
\resizebox{\hsize}{!}{\rotatebox{0}{\includegraphics{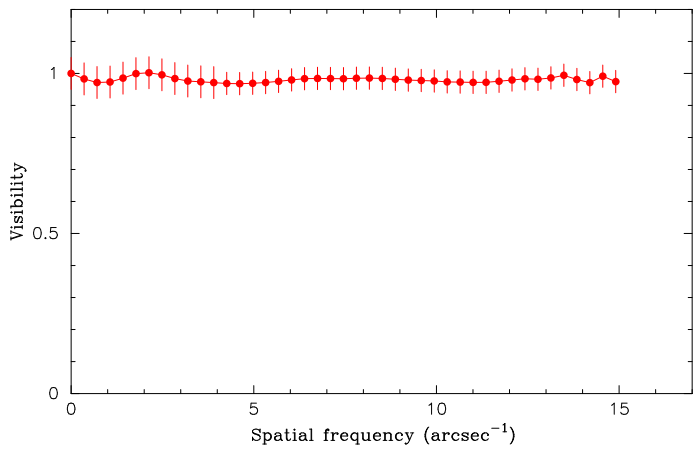}}}
%\resizebox{\hsize}{!}{\rotatebox{0}{\includegraphics{hr3126_naco_vis1D.ps}}}
\end{center}
\caption{
  Azimuthally averaged visibility of HR3126 measured at 2.24~\micron\
  with our NACO speckle interferometric observations. 
}
\label{hr3126_naco_vis1D}
\end{figure}

Figure~\ref{hr3126_naco_vis1D} shows 
the azimuthally averaged visibility of HR3126 extracted from our NACO speckle 
interferometric data taken at 2.24~\micron. 
The visibility is unity within the measurement errors up to the spatial
frequency of 15~arcsec$^{-1}$ that corresponds to the baseline of 7~m. 
The measured closure phase is also approximately zero within the covered
spatial frequencies. 
Therefore, 
HR3126 is unresolved with our NACO speckle interferometric observations
with the spatial resolution of 64~mas.

\subsection{VISIR mid-infrared imaging}
\label{appendix_res_visir}

As in the analysis of the SPHERE-ZIMPOL data above,
we obtained the FWHMs of the azimuthally averaged radial intensity profiles
of HR3126 and the PSF reference Canopus. 
The results are listed in Table~\ref{res_visir_table}. 
The FWHMs of HR3126 are nearly the same as those of Canopus, showing 
that HR3126 is unresolved from 7.9 to 19.5~\micron\ with spatial resolutions 
of 0\farcs4--0\farcs6. 
The FWHMs of Canopus with the J7.9 and SIV\_1 filters are slightly larger
than those of HR3126 due to variations in the atmospheric conditions.

%\clearpage
\section{Comparison of the interferometric observables between the 
AMBER+GRAVITY data and the radiative transfer model of the dust disk.
}

\begin{figure*}
\begin{center}
\resizebox{\hsize}{!}{\rotatebox{-90}{\includegraphics{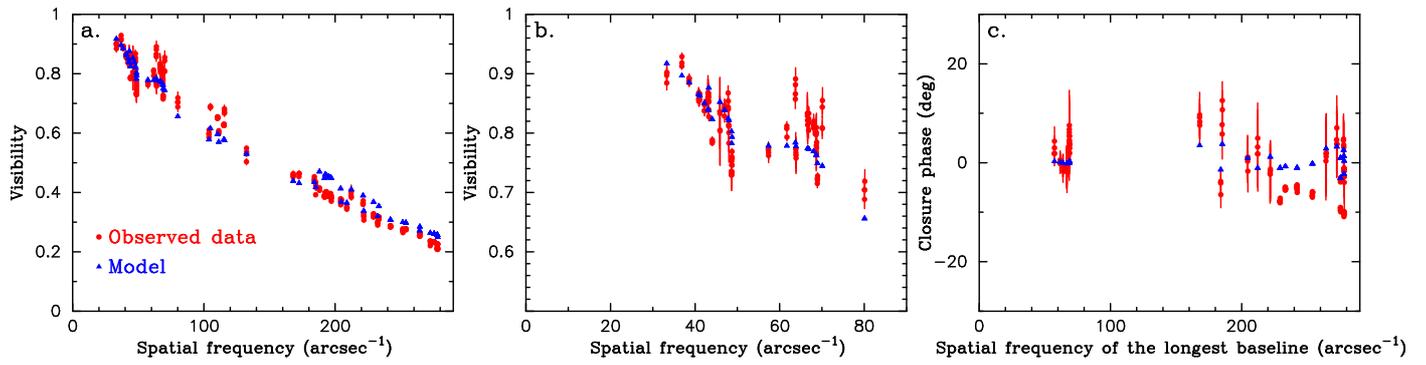}}}
%\resizebox{\hsize}{!}{\rotatebox{-90}{\includegraphics{fit_interf_model_3x1.ps}}}
\end{center}
\caption{
  Comparison of the interferometric observables observed at 2.25~\micron\ 
  with those
  predicted by the best-fit 2-D radiative transfer model in the continuum. 
Panels {\bf a}, {\bf b}, and {\bf c} show the comparison of the 
visibility, visibility at spatial frequencies lower than 90~arcsec$^{-1}$ 
(corresponding to baselines shorter than 42~m),
and closure phase, respectively. In each panel, 
the observed data and the values predicted by the best-fit model 
are shown by the red symbols with the error bars and the blue symbols, 
respectively. 
}
\label{fit_interf_model}
\end{figure*}

Figure~\ref{fit_interf_model} shows a comparison of the measured
interferometric observables (visibility and closure phase) and those
predicted by the best-fit dust disk model.

\clearpage
\section{Log of observations with VLTI/AMBER+GRAVITY, VLT/NACO,
  SPHERE-ZIMPOL, and VISIR}

Logs of our observations with VLTI/AMBER and GRAVITY, 
SPHERE-ZIMPOL, VLT/NACO, and VISIR are given in Tables.~\ref{obs_vlti_log},
\ref{obs_sphere_log}, \ref{obs_naco_log}, and \ref{obs_visir_log},
respectively.

\begin{table*}
\caption {
Summary of our VLTI/AMBER and GRAVITY observations of HR3126.
}
\begin{center}

\begin{tabular}{l c c c c c r l }\hline
\# & $t_{\rm obs}$ & Config. & $B_{\rm p}$ & PA     & Seeing   & $\tau_0$ &
${\rm DIT}\times{\rm N}_{\rm f}\times{\rm N}_{\rm exp}$ \\ 
& year/month/day UTC       &  & (m)       & (\degr) & (\arcsec)       &  (ms)    &  (sec)  \\
\hline
\multicolumn{8}{c}{AMBER}\\
\hline
1 & 2016/11/14 07:14:50 & A0-C1-D0 & 19.98/20.45/31.99 & 90/14/51    & 0.49 & 4.4 & $0.2\times200\times5$ \\
2 & 2016/11/14 08:15:41 & A0-C1-D0 & 21.23/20.01/31.88 & 102/24/64   & 0.90 &
2.9 & $0.2\times200\times2$ \\
3 & 2016/11/14 08:18:55 & A0-C1-D0 & 21.29/19.98/31.86 & 103/24/65   & 0.98 & 2.6 & $0.2\times200\times1$ \\
4 & 2016/12/19 07:13:06 & A0-C1-D0 & 22.19/19.08/31.08 & 118/35/80   & 0.72 & 8.2 & $0.2\times200\times5$ \\
5 & 2017/02/24 02:53:50 & A0-C1-D0 & 22.23/19.00/30.95 & 119/36/81   & 0.77 & 8.4 & $0.2\times200\times5$ \\
6 & 2017/02/24 03:49:19 & A0-C1-D0 & 22.54/17.97/29.71 & 130/44/93   & 0.62 & 11.6 & $0.2\times200\times5$ \\
7 & 2017/02/25 02:04:25 & A0-C1-D0 & 21.75/19.64/31.60 & 109/29/71   & 0.69 & 10.2 & $0.2\times200\times3$ \\
8 & 2017/02/25 05:17:03 & A0-C1-D0 & 22.59/15.59/26.77 & 150/59/114  & 0.68 & 6.2  & $0.2\times200\times4$ \\
9 & 2017/03/11 03:24:17 & A0-C1-D0 & 22.61/17.19/28.73 & 137/48/100  & 0.36 & 18.9 & $0.2\times200\times5$ \\
10 & 2016/11/19 07:06:47 & D0-G2-J3 & 37.20/61.51/98.57 & 24/30/27  & 0.69 & 2.6 & $0.2\times200\times4$ \\
11 & 2016/12/22 06:11:28 & D0-K0-J3 & 95.04/48.06/94.96 & 70/$-35$/40  & 0.75 & 3.8 & $0.2\times200\times5$ \\
12 & 2017/04/24 00:36:01 & D0-K0-J3 & 85.71/51.60/80.46 & 101/$-12$/65  & 0.75 & 3.1 & $0.2\times200\times5$ \\
13 & 2017/04/24 23:44:45 & D0-G2-J3 & 32.65/53.85/86.29 & 52/61/57  & 0.46 & 5.8 & $0.2\times200\times5$ \\
14 & 2017/04/25 00:46:51 & D0-G2-J3 & 29.73/48.92/78.40 & 62/71/68  & 0.51 &
5.6 & $0.2\times200\times5$ \\
15 & 2017/03/07 03:53:43 & A0-G1-J3 & 90.50/102.8/103.3 & 140/24/76 & 0.60 &
8.3 & $0.2\times200\times4$ \\
16 & 2017/03/15 00:52:27 & A0-J2-J3 & 129.0/87.11/122.9 & 89/$-25$/49 & 0.60 &
8.9 & $0.2\times200\times4$ \\
17 & 2017/03/15 01:49:58 & A0-J2-J3 & 128.9/89.59/117.2 & 102/$-17$/59 & 0.48
& 12.3 & $0.2\times200\times6$ \\
18 & 2017/03/15 02:56:02 & A0-J2-J3 & 126.4/91.15/107.8 & 116/$-7$/71 & 0.45 &
9.5 & $0.2\times200\times1$ \\
19 & 2017/03/15 03:51:00 & A0-J2-J3 & 122.9/91.48/97.89 & 129/1/82 & 0.55 &
6.1 & $0.2\times200\times7$ \\
\hline
\multicolumn{8}{c}{GRAVITY}\\
\hline
20 & 2018/12/19 07:19:10 & A0-G1-J2-J3 & 89.30/108.25/117.82/ & 162/$-168$/$-122$/
& 0.78 & 5.0 & $10 \times 20 \times 2$ \\
   &                     &             & 53.48/128.99/89.08 & $-113$/$-80$/$-61$
&     &     &    \\
21 & 2018/12/19 08:01:22 & A0-G1-J2-J3 & 90.53/106.37/112.36/ & 169/$-163$/$-114$/
& 0.49 & 7.0 & $10 \times 20 \times 2$ \\
   &                     &             & 51.21/127.80/90.05 & $-105$/$-71$/$-52$
&     &     &    \\
\hline
\label{obs_vlti_log}
\vspace*{-7mm}

\end{tabular}
\end{center}
\tablefoot{
$B_{\rm p}$: Projected baseline length.  PA: Position angle of the baseline 
vector projected onto the sky. 
DIT: Detector Integration Time.  $N_{\rm f}$: Number of frames in each 
exposure.  $N_{\rm exp}$: Number of exposures. 
The seeing and the coherence time ($\tau_0$) were measured in the visible. 
}
\end{table*}

\clearpage
\begin{table*}
\caption {
Summary of our VLT/SPHERE-ZIMPOL observations of HR3126 and the calibrator
HD65273 with the V filter. 
}
\begin{center}

\begin{tabular}{l l c c c l c c c}\hline
\# & Object & $t_{\rm obs}$ & Seeing   & $\tau_0$ &
${\rm DIT}\times{\rm N}_{\rm f}\times{\rm N}_{\rm exp}$&Strehl ($H$) & 
Strehl ($V$) & FWHM ($V$)\\ 
   &        & (UTC)    & (\arcsec) &  (ms)    &  (sec)  & & & (mas) \\
\hline
\multicolumn{9}{c}{2016 March 15 (UTC)}\\
\hline
1 & HR3126  & 00:58:50 & 0.89 & 3.9 & $10\times10\times2$ & 0.90 & 0.39 & 22.6\\
2 & HD65273 & 01:42:17 & 0.75 & 3.9 & $10\times10\times2$ & 0.88 & 0.31 & 25.1\\
\hline
\label{obs_sphere_log}
\vspace*{-7mm}

\end{tabular}
\end{center}
\tablefoot{
DIT: Detector Integration Time.  ${\rm N}_{\rm f}$: Number of frames in each 
exposure. ${\rm N}_{\rm exp}$: Number of exposures for each polarization 
component. 
The seeing and the coherence time ($\tau_0$) were measured in the visible. 
}
\end{table*}

\begin{table}
\caption {
Summary of our VLT/NACO observations of HR3126 and the calibrator
$\beta$~Vol with the IB2.24 filter.
}
\begin{center}

\begin{tabular}{l l c c c l}\hline
\# & Object & $t_{\rm obs}$ & Seeing   & $\tau_0$ &
${\rm DIT}\times{\rm N}_{\rm f}$ \\ 
   &        & (UTC)    & (\arcsec) &  (ms)    &  (ms)  \\
\hline
\multicolumn{6}{c}{2017 April 10 (UTC)}\\
\hline
1 & HR3126 & 23:48:16 & 1.09 & 5.8 &  $150\times1002$\\
2 & HR3126 & 23:57:38 & 1.15 & 5.6 &  $150\times925$\\
\hline
\multicolumn{6}{c}{2017 April 11 (UTC)}\\
\hline
3 & HR3126 & 00:15:26 & 0.81 & 6.2 & $150\times943$ \\
4 & HR3126 & 00:29:17 & 0.76 & 6.6 & $150\times901$ \\
5 & HR3126 & 00:48:32 & 0.67 & 7.1 & $150\times933$ \\
6 & $\beta$~Vol & 00:57:02 & 0.66 & 6.9 & $150\times911$ \\
7 & HR3126 & 01:03:36 & 0.70 & 6.1 & $150\times904$ \\
8 & $\beta$~Vol & 01:10:38 & 0.66 & 6.9 & $150\times911$ \\
9 & HR3126 & 01:17:01 & 0.79 & 5.4 & $150\times905$ \\
10 & $\beta$~Vol & 01:23:58 & 1.20 & 3.5 & $150\times907$ \\
\hline
\label{obs_naco_log}
\vspace*{-7mm}

\end{tabular}
\end{center}
\tablefoot{
DIT: Detector Integration Time.  $N_{\rm f}$: Number of frames. 
The seeing and the coherence time ($\tau_0$) were measured in the visible. 
}
\end{table}

\begin{table*}
\caption {
Summary of our VLT/VISIR observations of HR3126 and the calibrator 
Canopus.
}
\begin{center}

\begin{tabular}{l l c c c c c c}\hline
\# & Object & $t_{\rm obs}$ & Filter & Seeing   & PWV & $t_{\rm int}$ 
&$N_{\rm nod}$  \\ 
   &        & (UTC)       &        & (\arcsec)&(mm) &  (sec)  & \\
\hline
\multicolumn{8}{c}{2015 November 1 (UTC)}\\
\hline
1 & Canopus  & 07:15:07 & Q1 & 1.1 & 1.6 & 575.189  & 5 \\
2 & Canopus  & 07:28:27 & Q3 & 1.1 & 1.7 & 618.839  & 5 \\
3 & HR3126   & 07:49:09 & Q1 & 1.3 & 1.7 & 575.189  & 5 \\
4 & HR3126   & 08:01:57 & Q3 & 1.4 & 1.7 & 618.839  & 5 \\
5 & Canopus  & 08:22:53 & J7.9 & 1.3 & 1.8 & 502.750 & 2 \\
6 & Canopus  & 08:32:39 & SIV\_1 & 1.1 & 1.8 & 472.894 & 2 \\
7 & HR3126   & 08:49:23 & J7.9 & 1.2 & 1.8 & 509.134 & 2 \\
8 & HR3126   & 08:59:55 & SIV\_1 & 1.2 & 1.8 & 472.894 & 2\\
\hline
\multicolumn{8}{c}{2015 November 2 (UTC)}\\
\hline
9 & Canopus  & 06:15:15 & NeII & 1.1 & 2.8 & 480.016 & 2 \\
10 & HR3126  & 06:29:24 & NeII & 1.2 & 2.9 & 480.016 & 2 \\
\hline
\label{obs_visir_log}
\vspace*{-7mm}

\end{tabular}
\end{center}
\tablefoot{
PWV: Precipitable Water Vapor. $t_{\rm int}$: Total integration time. 
$N_{\rm nod}$: Number of nodding cycles. 
The seeing was measured in the visible. 
}
\end{table*}

\end{document}